\documentclass[twocolumn,showpacs,nobibnotes,nofootinbib,notitlepage,aps,prd,longbibliography,superscriptaddress]{revtex4-1}
\usepackage{amsmath,amssymb}
\usepackage[dvipsnames]{xcolor}
\usepackage[colorlinks,linkcolor=Maroon,anchorcolor=blue,citecolor=MidnightBlue]{hyperref} 
\usepackage{graphicx}
\usepackage{slashed}
\usepackage{ulem}
\usepackage{bm}
\usepackage{mathrsfs}
\usepackage{array}
\usepackage{comment}
\usepackage{multirow}

\newcommand{\figref}[1]{Fig.~\ref{#1}}
\newcommand{\rcite}[1]{Ref.~\cite{#1}}
\newcommand{\eref}[1]{Eq.~\eqref{#1}}

\allowdisplaybreaks

\begin{document}

\author{Yin-Da Guo}
\email{yinda.guo@mail.sdu.edu.cn}
\affiliation{Key Laboratory of Particle Physics and Particle Irradiation (Ministry of Education),\\Institute of Frontier and Interdisciplinary Science, \\Shandong University, Qingdao 266237, China}

\author{Nayun Jia}
\email{nayun.jia@foxmail.com}
\affiliation{Department of Physics, Southern University of Science and Technology, Shenzhen 518055, China}
\affiliation{Key Laboratory of Cosmology and Astrophysics (Liaoning) \& College of Sciences, Northeastern University, Shenyang 110819, China}

\author{Shou-Shan Bao}
\email{ssbao@sdu.edu.cn}
\author{Hong Zhang}
\email{hong.zhang@sdu.edu.cn}
\affiliation{Key Laboratory of Particle Physics and Particle Irradiation (Ministry of Education),\\Institute of Frontier and Interdisciplinary Science, \\Shandong University, Qingdao 266237, China}

\author{Xin Zhang}
\email{zhangxin@mail.neu.edu.cn}
\affiliation{Key Laboratory of Cosmology and Astrophysics (Liaoning) \& College of Sciences, Northeastern University, Shenyang 110819, China}
\affiliation{Key Laboratory of Data Analytics and Optimization for Smart Industry (Ministry of Education), Northeastern University, Shenyang 110819, China}
\affiliation{National Frontiers Science Center for Industrial Intelligence and Systems Optimization, Northeastern University, Shenyang 110819, China}

\date{\today}

\title{Evolution and detection of vector superradiant instabilities}

\begin{abstract}
Ultralight vectors can extract energy and angular momentum from a Kerr black hole (BH) due to superradiant instability, resulting in the formation of a BH-condensate system. In this work, we carefully investigate the evolution of this system numerically with multiple superradiant modes. Simple formulas are obtained to estimate important timescales, maximum masses of different modes, as well as the BH mass and spin at various times. Due to the coexistence of modes with small frequency differences, the BH-condensate system emits gravitational waves with a unique beat signature, which could be directly observed by current and projected interferometers. Besides, the current BH spin-mass data from the binary BH merger events already exclude the vector mass in the range $5\times 10^{-15}\ \mathrm{eV}  <\mu< 9\times 10^{-12}\ \mathrm{eV}$.
\end{abstract}
%
\maketitle
\pagebreak

\section{Introduction}

The rotational energy of a Kerr black hole (BH) can be extracted by particle fissioning in the ergosphere, the so-called Penrose process \cite{Penrose:1971uk}. It is later generalized to classical waves \cite{Misner:1972kx} and is named ``superradiant scattering" \cite{Press:1972zz}.  The superradiant process occurs when the wave frequency $\omega$ satisfies $\omega < m\Omega_H$, where $m$ is the magnetic number and $\Omega_H$ is the angular velocity at the outer horizon. If there exists something like a mirror far from the BH reflecting the outgoing wave, the superradiant scattering could happen repeatedly, causing an exponential growth of the wave energy. In particular, an ultralight bosonic field can form bound states within the gravitational potential of the host BH, which prevents the bosons from escaping to infinity.  The bound states then continuously extract the energy and angular momentum from the host BH, forming a large-size BH-condensate system, which could be observed with optical telescopes or gravitational wave (GW) interferometers. For a comprehensive review of existing studies, we refer the readers to \rcite{Brito:2015oca}.

To detect the BH-condensate system, both direct and indirect methods are available. In terms of direct detection, the authors of Refs.~\cite{Yoshino:2014wwa,Isi:2018pzk,Sun:2019mqb} detail a targeted search for GWs using data from the (Advanced) Laser Interferometer Gravitational Wave Observatory (LIGO) \cite{LIGOScientific:2014pky}. To distinguish the system from other monochromatic GW sources, such as neutron stars, it has been suggested that the GWs from the system have a positive frequency drift due to self-gravity, while the GWs from neutron stars have a frequency drift in the opposite direction \cite{Arvanitaki:2014wva,Baryakhtar:2017ngi}. In a previous calculation with ultralight scalars \cite{Guo:2022mpr}, we have shown that the subdominant modes with overtone $n>0$ could coexist for a long time with the dominant mode. With slightly different frequencies, the interference of these modes could produce a unique beat signature of the emitted GW. The beat period ranges from milliseconds to months, depending on the mass of the host BH. Explicit calculation also shows that the beat strength is within the capability of current and projected interferometers. Thus the beat signature is another fingerprint of the BH-condensate systems. In addition, unresolved GWs from BH-condensate systems can be studied as potential stochastic backgrounds \cite{Brito:2017wnc,Brito:2017zvb}. 

The existence of BH-condensate systems in our Universe could also be indirectly studied from the observed BH spin-mass distribution. Fast-spinning BHs would quickly lose their angular momenta to the condensates around them, leading to forbidden regions in the BH spin-mass plane \cite{Arvanitaki:2010sy}. Thus analyzing the BH data from merger events allows us to identify both favored and unfavored ranges of the boson mass \cite{Arvanitaki:2016qwi,Cardoso:2018tly,Ng:2019jsx,Ng:2020ruv,Fernandez:2019qbj,Cheng:2022jsw}. Then experiments can be designed to search for these ultralight bosons in the favored mass ranges.

The most widely-studied scenario is the ultralight scalar \cite{Zouros:1979iw,Cardoso:2005vk,Arvanitaki:2009fg,Arvanitaki:2010sy,Konoplya:2011qq,Witek:2012tr,Yoshino:2013ofa,Arvanitaki:2014wva,Brito:2014wla,Arvanitaki:2016qwi,Endlich:2016jgc,Brito:2017zvb,Brito:2017wnc,Ficarra:2018rfu,Ng:2019jsx,Fernandez:2019qbj,Sun:2019mqb,Ng:2020ruv,Roy:2021uye,Hui:2022sri}, with superradiance rate either from Detweiler's famous analytic approximation \cite{Detweiler:1980uk} or from numerical calculations \cite{Dolan:2007mj,Yoshino:2014wwa,Yoshino:2015nsa}. Nonetheless, the analytic approximation is inconsistent with the numerical results, with the error as large as $150\%$. The puzzle is solved by generalizing Detweiler's result to the next-to-leading order \cite{Bao:2022hew, Bao:2023xna}. The new approximation also has a compact form and reduces the error to $\lesssim 10\%$. It has been applied to study ultralight scalars as well as the GW signature of the BH-condensate systems \cite{Guo:2022mpr, Cheng:2022jsw}.

Ultralight vector fields, such as dark photons \cite{Holdom:1985ag,Essig:2013lka}, could also form such a BH-condensate system via superradiance \cite{East:2013mfa,Baryakhtar:2017ngi,East:2017mrj,East:2017ovw,Cardoso:2018tly,East:2018glu,Isi:2018pzk,Siemonsen:2019ebd,Chen:2022nbb,Jia:2023see}. Existing studies mainly focus on the most unstable mode with overtone number $n=0$. The GW signal from such a BH-condensate system is then monochromatic with approximately constant amplitude within its lifetime. If the duration is long, it would be indistinguishable from other monochromatic GW sources, such as neutron stars. Similar to the scalar case, the subdominant modes could be important to break the degeneracy. In this work, we apply the analysis in Refs.~\cite{Guo:2022mpr,Cheng:2022jsw} to ultralight vectors.  Particularly, we focus on the parameter space where the Compton wavelengths of the vectors are comparable to the BH radius, i.e., $G_\mathrm{N}M\mu/(\hbar c)\sim\mathcal{O}(1)$, where $M$ is the mass of the BH and $\mu$ is the vector mass. This region of parameters has the strongest superradiant instability and is most relevant in phenomenology.

To our best knowledge, the beat feature in the GW emitted by BH-vector-condensate systems is first studied in Ref.~\cite{Siemonsen:2019ebd}. The authors focus on the special ``overtone mixing'' regions where the large-overtone modes have a superradiance rate comparable to or even larger than the small-overtone modes. These modes are then similar in size in the evolution, leading to the beat feature in the radiated GW. In this work, we show the beat feature could always happen even when the two modes are very different in size. Specifically, we show the beat signal can reach 5\% even when the subdominant mode is only $10^{-3}$ of the dominant mode in mass.

This paper is organized as follows. In Sec.~\ref{sec:vecBH}, we briefly review the calculation of the superradiance rate and the GW emission rate of different modes. Both effects are important for the time evolution of the BH-condensate systems. In Sec.~\ref{sec:Single-mode}, we focus on the $n=0$ mode and discuss the effects of the initial parameters on the evolution. Then our attention shifts to the multi-mode case in Sec.~\ref{sec:Multi-mode}, where we divide the evolution into stages with different $j$. In Sec.~\ref{sec:detec}, we further explore the direct and indirect approaches to detect the vector superradiance, using GW signals and BH spin-mass distributions. Finally, we summarize in Sec.~\ref{sec:summary}.

Throughout the paper, we adopt the Planck units $G_\mathrm{N}=\hbar=c=1$, and the $\{+,-,-,-\}$ signature.

\section{Theoretical framework}
\label{sec:vecBH}

In this section, we first review the Proca field in Kerr spacetime. Then we discuss in detail the superradiance rate and the GW emission rate, which are the two essential pieces in the evolution of the BH-condensate systems. Similar to the case with a scalar field, the ultralight vector field influences modestly on the BH mass but greatly on the BH spin. At the end of this section, we obtain the evolution equations of the BH-condensate systems, which will be used in the rest of this paper.

\subsection{The Proca equation in Kerr spacetime}

The Kerr spacetime metric, characterized by  mass $M$ and angular momentum $J$, has the following form in the Boyer-Lindquist coordinates \cite{Boyer:1966qh}:
\begin{equation}
\begin{aligned}
ds^2=&\left( 1-\frac{2Mr}{\Sigma} \right) dt^2+\frac{4aMr}{\Sigma}\sin ^2\theta dtd\varphi -\frac{\Sigma}{\Delta}dr^2 \\
&-\Sigma d\theta ^2-\left[ \left( r^2+a^2 \right) \sin ^2\theta +2\frac{Mr}{\Sigma}a^2\sin ^4\theta \right] d\varphi ^2,
\label{eq:KerrMetric}
\end{aligned}
\end{equation}
where
\begin{align}
    a & \equiv J/M,\\
    \Delta & \equiv r^2-2 M r+a^2,\\
    \Sigma & \equiv r^2+a^2 \cos ^2 \theta.
\end{align}
The inner horizon $r_-$ and outer horizon $r_+$ are
\begin{equation}
    r_{ \pm}=M \pm \sqrt{M^2-a^2}.
\end{equation}
The variable $a$ denotes the angular momentum of the BH per unit mass. Throughout this paper, we use the dimensionless variable $a_*\equiv a/M$ and refer to it as the BH {\it spin} for brevity.

The Lagrangian of a free massive vector field $A_\mu$ is
\begin{equation}\label{eq:Lagrangian}
    \mathcal{L}=-\frac{1}{4} F_{\rho \sigma} F^{\rho \sigma}+\frac{1}{2} \mu^2 A_\lambda A^\lambda,
\end{equation}
where $\mu$ is the vector mass and
\begin{equation}\label{eq:field_strength_tens}
    F_{\rho \sigma} \equiv \nabla_\rho A_\sigma-\nabla_\sigma A_\rho
\end{equation}
is the field strength tensor. The Proca equation can be derived straightforwardly from the Lagrangian \cite{Proca:1936fbw},
\begin{equation}
    \nabla_\rho F^{\rho \sigma}+\mu^2 A^\sigma=0,
    \label{eq:Proca}
\end{equation}
in which the Lorenz gauge condition,
\begin{equation}\label{eq:Lorenz_Gauge}
    \nabla_\rho A^\rho = 0,
\end{equation}
is automatically satisfied. The corresponding stress-energy tensor of the vector field is
\begin{align}
    \begin{split}
        T_{\rho\sigma}= & F_{\rho\beta}F_{\sigma}{}^{\beta}-\mu^2A_{\rho}A_{\sigma}\\
        & \hspace{1cm} +g_{\rho\sigma}\left(-\frac{1}{4}F_{\alpha\beta}F^{\alpha\beta}+\frac{1}{2}\mu^2A_{\alpha}A^{\alpha}\right).
    \end{split}
\end{align}

The Proca equation \eqref{eq:Proca} has quasi-bound solutions with complex eigenfrequency $\omega+i\Gamma$. Calculating these quasi-bound states in a general case turns out to be quite difficult. Nevertheless, the system exhibits hydrogen-like solutions in the non-relativistic limit, allowing us to write the vector field as \cite{Baryakhtar:2017ngi},
\begin{equation}
    A^{\rho}(t, r, \theta, \varphi)=\frac{1}{\sqrt{2 \omega}}\left[ \Psi^\rho(r,\theta, \varphi)e^{-i \omega t}+\text { c.c. }\right],
    \label{eq:vector_function_Single}
  \end{equation}
where we have ignored the imaginary part of the eigenfrequency $i\Gamma$ since it is much smaller than the real part $\omega$. This point will be evident below by comparing \eref{eq:omega} to \eref{eq:SR_rate_Baumann}. With the Lorenz gauge condition, the time component of $A^\rho$ can be solved from its spatial components. The variables of the spatial components can be separated as
\begin{equation}\label{eq:VarSepPsi}
    \mathbf{\Psi}(r,\theta, \varphi)=R_{n l}(r) \mathbf{Y}_{l, j m}(\theta, \varphi),
    \end{equation}
where the integers $n,\ j,\ l$ and $m$ represent the overtone number, the total angular momentum number, the azimuthal number, and the magnetic number, respectively. Another widely used quantity is the principal number $\bar{n}$, defined as $n+l+1$. In the rest of the paper, we will refer to a specific mode using a generalized spectroscopic notation $^{2s+1}L_{j,m}^n$, with $L=S,P,D,F\dots$ for $l=0,1,2,3$, respectively. When $l$ is undetermined, the notation $^{2s+1}l_{j,m}^n$ is also used. In the presence of multiple modes, the vector field \eqref{eq:vector_function_Single} can be expressed as
\begin{equation}
    A^{\rho}=\sum_{nljm}\sqrt{\frac{N_{nljm}}{2 \omega_{nljm}}}\left(\mathcal{A}^{\rho}_{(nljm)}+\mathcal{A}^{*\rho}_{(nljm)}\right), 
    \label{eq:A_multi} 
  \end{equation}
where $\mathcal{A}_{(nljm)}^\rho\equiv\Psi_{(nljm)}^\rho e^{-i \omega_{nljm} t}$ and $N_{nljm}$ is the occupation number of the $^{3}l_{j,m}^n$ mode. In this definition, the field $\mathcal{A}^\rho_{(nljm)}$ is normalized as a single-particle wavefunction. 

Inserting Eq.~\eqref{eq:VarSepPsi} into the equation of motion in Eq.~\eqref{eq:Proca} and taking the non-relativistic limit, one could solve for the radial and angular pieces. The radial wavefunction $R_{n l}(r)$ can be expressed in terms of Laguerre polynomials \cite{Brito:2014wla},
\begin{equation}
    R_{nl}(\tilde{r})=\sqrt{\left(\frac{2}{\bar{n}r_{a}}\right)^3\frac{n!}{2\bar{n}(\bar{n}+l)!}}\mathrm{e}^{-\tilde{r}/2}\tilde{r}^l L_{n}^{2l+1}(\tilde{r}),
\end{equation}
with $\tilde{r} \equiv 2r/(\bar{n}r_\mathrm{B})$, and $r_\mathrm{B}\equiv 1/(M\mu^2)$ is analogous to the Bohr radius in the hydrogen atom. The angular piece $\mathbf{Y}_{l, j m}(\theta, \varphi)$ in Eq.~\eqref{eq:VarSepPsi} is the pure-orbital vector spherical harmonic function,  which satisfies the eigenequation \cite{Thorne:1980ru,Baryakhtar:2017ngi},
\begin{equation}
    -r^{2} \nabla^{2} \mathbf{Y}_{l, j m}=l(l+1) \mathbf{Y}_{l, j m}.
\end{equation}
Interestingly, the eigenenergy in the $M\mu\ll 1$ limit is similar to the hydrogen case \cite{Baryakhtar:2017ngi},
\begin{equation}
    \omega_{nljm}\approx\mu\left[1-\frac{(M\mu)^2}{2\bar{n}^2}\right].
    \label{eq:Approxomega} 
\end{equation}
Below we will refer to $M\mu$ as {\it mass coupling} for brevity.

The above discussion assumes the non-relativistic limit. In the relativistic case, one could not simply separate the variables as in Eq.~\eqref{eq:VarSepPsi} \cite{Brito:2015oca}. Previous efforts include a semi-analytical treatment under slow-rotation approximation \cite{Pani:2012bp,Pani:2012vp,Pani:2013pma} and a numerical approach pioneered by \rcite{Cardoso:2018tly}, which obtains precise numerical results of the $^{3}S_{1,1}^0$ modes. In 2018, inspired by the work of \rcite{Lunin:2017drx}, the authors of \rcite{Frolov:2018ezx} successfully separated the variables in the Proca equation with a general Kerr-NUT-(A)dS BH metric in all dimensions. Shortly after that, the obtained coupled equations in four-dimensional Kerr metric are solved both numerically and analytically \cite{Dolan:2018dqv,Baumann:2019eav}. Central to the variable-separating process is the ansatz
\begin{equation}
    A^\rho=B^{\rho \sigma} \nabla_\sigma Z,
\end{equation}
with
\begin{equation}
    Z(t,r,\theta,\varphi)=e^{-i (\omega+i\Gamma) t+i m \varphi} \mathcal{R}(r) \mathcal{S}(\theta),
\end{equation}
where $\mathcal{R}$ and $\mathcal{S}$ are the radial and angular functions, respectively. The polarization tensor $B^{\mu \nu}$ satisfies
\begin{equation}
    B^{\rho \lambda}\left(g_{\lambda \sigma}+i \nu h_{\lambda \sigma}\right)=\delta^\rho{}_\sigma,
\end{equation}
where $\nu$ is a complex constant to be determined and $h_{\lambda \sigma}$ is the principal tensor \cite{Frolov:2017kze}. More details of the calculation can be found in Refs.~\cite{Lunin:2017drx,Frolov:2017kze,Frolov:2018pys,Frolov:2018ezx,Dolan:2018dqv,Baumann:2019eav,Lunin:2019pwz}.

\subsection{Superradiant rates}\label{subsec:srrate}

In this work, we employ the analytical results in the $M\mu\ll1$ limit, first presented in \rcite{Baumann:2019eav}. The real part of the eigenfrequency with high-order correction is
\begin{align}\label{eq:omega}
\begin{split}
\frac{\omega_{n l j m}}{\mu}&=1-\frac{\alpha^2}{2 \bar{n}^2}-\frac{\alpha^4}{8 \bar{n}^4}+\frac{f_{\bar{n} l j}}{\bar{n}^3} \alpha^4+\frac{h_{l j}}{\bar{n}^3} a_* m \alpha^5+\cdots,
\end{split}
\end{align}
where $\bar{n}=n+l+1$, $\alpha \equiv M\mu$ and
\begin{align}
    f_{\bar{n} l j} & = -\frac{4(6 l j+3 l+3 j+2)}{(l+j)(l+j+1)(l+j+2)}+\frac{2}{\bar{n}}, \\
    h_{l j} & = \frac{16}{(l+j)(l+j+1)(l+j+2)}.
\end{align}
The imaginary part, which is also called the superradiance rate, has the following form:
\begin{equation}\label{eq:SR_rate_Baumann}
    \Gamma _{nljm}=  2\left( r_+/M \right) C_{nlj}g_{jm}\left( m\Omega_H-\omega_{nljm} \right) (M\mu) ^{2l+2j+5},
\end{equation}
where
\begin{align}
	C_{nlj}& =  \frac{2^{2l+2j+1}(n+2l+1)!}{\left( n+l+1 \right) ^{2l+4}n!}\left[ \frac{(l)!}{(l+j)!(l+j+1)!} \right] ^2 \nonumber\\
	&\hspace{1cm}\times \left[ 1+\frac{2(1+l-j)(1-l+j)}{l+j} \right] ^2,\\
    g_{j m}& = \prod_{k=1}^j{\left[ k^2\left( 1-{a_*}^2 \right) +\left( a_*m-2r_+\omega_{nljm} \right) ^2 \right]},
\end{align}
and $\Omega_H\equiv{a}/{\left(2 M r_{+}\right)}$ is the angular velocity at the outer horizon.

The superradiance rate in Eq.~\eqref{eq:SR_rate_Baumann} depends on $M\mu$, $a_*$ as well as $n,l,j$ and $m$. It is crucial to understand which modes are the most important for a given value of $M\mu$ and $a_*$. Fixing the total angular momentum $j$, the value of $m$ varies from $-j$ to $j$. The value of $l$ can be $j-1$, $j$, and $j+1$ for vector fields. The overtone number $n$ can be any non-negative integer. From the superradiance condition $\omega<m\Omega_H$, it is natural to conjecture that the mode with the same $j$ but a smaller value of $m$ has a slower superradiance rate. Indeed, we find the rate of the $m=j$ mode is larger than the $m<j$ modes by at least 7 orders of magnitude. Thus the modes with $m<j$ are never important in phenomenology. 

\begin{figure}[htbp]
    \centering
    \includegraphics[width=0.48\textwidth]{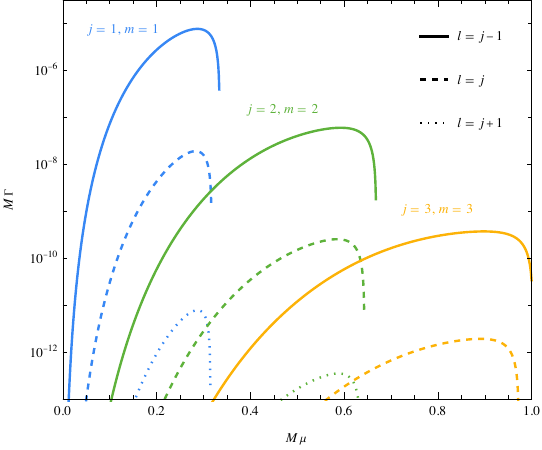}
    \caption{
    The superradiance rate $\Gamma$ as a function of the mass coupling $M\mu$, calculated with the analytical approximation in \eref{eq:SR_rate_Baumann}. The BH spin $a_*$ is chosen as 0.9, and the overtone number $n$ is 0. The curves with different colors correspond to different values of $j$ and $m$. For each set of $j$ and $m$, three modes with $l=j-1$ (solid), $l=j$ (dashed), and $l=j+1$ (dotted) are plotted and compared.}
    \label{fig:SR_rate_diff_modes}
\end{figure}

Utilizing the analytical result in Eq.~\eqref{eq:SR_rate_Baumann}, we plot the dependence of the superradiance rate on the mass coupling $M\mu$ in \figref{fig:SR_rate_diff_modes}, with BH spin $a_*=0.9$, $m=j$ and $n=0$ for illustration. Curves with different values of $l$ are compared. All curves have the same qualitative behavior. Each of them first rises with $M\mu$ and drops rapidly to below zero after reaching the maximum. The fast dropping edge is a consequence of the factor $(m\Omega_H-\omega_{nljm})$ in Eq.~\eqref{eq:SR_rate_Baumann}. For a given value of $M\mu$, the mode with $l=j-1$ has the largest rate, followed by the $l=j$ mode with a rate roughly two decades smaller. The rate of the  $l=j+1$ mode is suppressed by another two orders of magnitude. We thus conclude that a vector field with both orbital angular momentum and spin aligned with the BH spin, i.e. the $^{3}(j-1)^n_{j,j}$ has the largest superradiance rate. Below we refer to these modes as {\it doubly-aligned modes}.

\begin{figure}[htbp]
    \centering
    \includegraphics[width=0.48\textwidth]{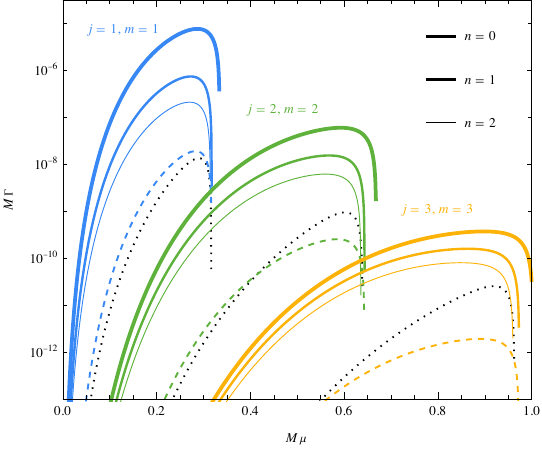}
    \caption{The superradiance rate $\Gamma$ as a function of the mass coupling $M\mu$, calculated with the analytical approximation in \eref{eq:SR_rate_Baumann}. The BH spin $a_*$ is chosen as 0.9. The curves with different colors correspond to different values of $j$ and $m$. The superradiance rates of the $^3(j-1)_{j,j}^n$ modes with $n=0,1,2$ are displayed by solid curves with decreasing thickness. For comparison, the superradiance rates of $^3j_{j,j}^0$ mode are shown by the dashed curves. The scalar superradiance rates in modes $^1j_{j,j}^0$ are shown by the dotted curves for comparison.}
    \label{fig:SR_rate_diff_n}
\end{figure}

The doubly-aligned modes are labelled by two numbers $j$ and $n$. In Fig.~\ref{fig:SR_rate_diff_n}, we further compare the superradiance rates of these modes with the same $j$ but different $n$. The superradiance rate is smaller with a larger value of $n$. Nonetheless, the suppression with different $n$ is much milder than those from varying $m$ and $l$. In Fig.~\ref{fig:SR_rate_diff_n}, the curves for the $^3j_{j,j}^0$ modes are also shown. Numerical calculation shows they are comparable to the doubly-aligned modes $^3(j-1)_{j,j}^n$ with $n=5,9,14$ for $j=1,2,3$, respectively. For a given value of $M\mu$, it is clear that the {\it dominant mode} is $^3(j-1)_{j,j}^0$ and the {\it subdominant mode} is $^3(j-1)_{j,j}^1$, where $j$ is the smallest positive integer satisfying $j>\omega_{0(j-1)jj}/\Omega_H$.

The subdominant modes are important phenomenologically. In the case of a scalar field, it has been shown that the subdominant modes can coexist with the dominant ones for a very long time. Due to the tiny difference in their frequencies, the presence of the subdominant modes results in an observable modulation of the GW emission from the BH-scalar-condensate systems~\cite{Guo:2022mpr}. The period of the modulation ranges from milliseconds to months, depending on the BH mass. This unique signal has been proposed to search for the ultralight scalar field experimentally. In this work, we continue to study the effect of the subdominant modes on the evolution of a BH-vector-condensate system. We show that similar modulation in the GW emission is also present, which could be utilized to search for ultralight vectors.

Although not closely related to the calculation of this work, it is interesting to compare the superradiance rate of a vector field to that of a scalar field. The dominant modes $^1j_{j,j}^0$ of a scalar field are represented with dotted black curves in Fig.~\ref{fig:SR_rate_diff_n}. The curves have the same qualitative behavior as those of the vector fields with the same $j$. Especially, their zero points are approximately at the same location, which is determined by the superradiance condition $\omega<m\Omega_H$. The size of the scalar superradiance rate is at least one decade smaller than the fastest vector superradiance rate. Numerical comparison shows the fastest scalar superradiance rate is roughly comparable to the vector doubly-aligned modes with the same $j$ but $n=5$. Thus if a vector has the same mass as a scalar, the former would extract the BH spin so efficiently that we are not able to observe the effect of the latter via BH superradiance processes. 

\begin{figure}[htbp]
    \centering
    \includegraphics[width=0.48\textwidth]{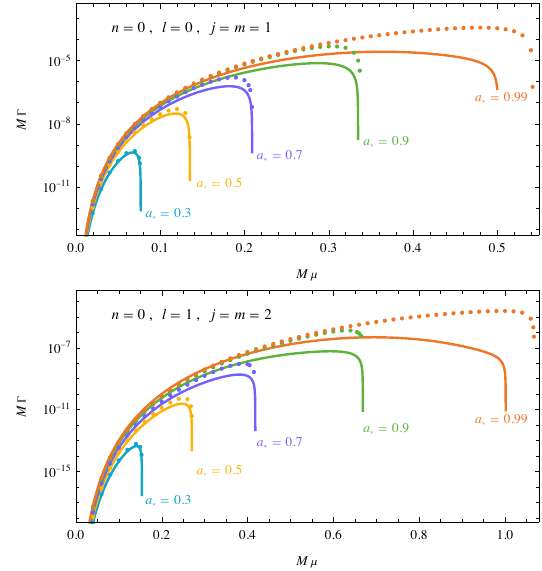}
    \caption{The superradiance rate $\Gamma$ as a function of the mass coupling $M\mu$ for the $^3S_{1,1}^0$ mode (upper panel) and the $^3P_{2,2}^0$ mode (lower panel). The curves of different colors represent different values of BH spin $a_*$. The solid curves are from the analytic approximation in Eq.~\eqref{eq:SR_rate_Baumann}, while the dotted curves are from numerical calculation using the method in Ref.~\cite{Dolan:2018dqv}. }
    \label{fig:SR_rate_diff_aStar}
\end{figure}

Figure~\ref{fig:SR_rate_diff_aStar} further shows the dependence of the superradiance rate of the $^3S_{1,1}^0$ and $^3P_{2,2}^0$ modes on BH spin $a_*$. With $a_*$ decreasing, the superradiance rate decreases, and the $M\mu$ range in which the rate is positive shrinks. For each value of $M\mu$, there is a critical BH spin below which the superradiance could not happen. In Fig.~\ref{fig:SR_rate_diff_aStar} we also compare the analytic approximation in Eq.~\eqref{eq:SR_rate_Baumann} to the numerical results using the direct integration method in Ref.~\cite{Dolan:2018dqv}. It is expected that the error of \eref{eq:SR_rate_Baumann} becomes larger at larger $M\mu$, since the analytical expression is effectively a Taylor expansion of $M\mu$. To have a sense of the difference, the relative error for the $^3S_{1,1}^0$ mode with $a_*=0.99$ is approximately $2.23\%$ at $M\mu=0.1$ and grows to roughly $90.0\%$ as $M\mu=0.4$.

\subsection{GW emission fluxes}\label{subsec:GW_emission_rates}

Rotating vector-condensate around the host BH emits GWs. In this part, we first focus on the GW emission by the dominant and subdominant modes, which are important in the evolution of BH-condensate systems. Then we move on to the interference of these modes, which produces the unique beat signals.

In the case with only a single mode $^3l_{j,m}^n$, we follow the method described in Ref.~\cite{Brito:2014wla} and calculate the GW emission flux in the limit of $M\mu\ll1$ using the Newman-Penrose (NP) formalism. An additional assumption of flat background metric is made to simplify the algebra. The GW emission flux of the $^3l_{j,m}^n$ mode at the leading order of $M\mu$ can be expressed as
\begin{align}
    \dot{E}^{(nljm)}_\mathrm{GW} \approx C^{(nljm)}_\mathrm{GW} \left(\frac{M_\mathrm{v}^{(nljm)}}{M}\right)^2(M\mu)^{4l+10}.
    \label{eq:GW_single}
\end{align}
For the dominant and subdominant modes with $j=1,2$, the corresponding coefficients are
\begin{subequations}\label{eq:GW_single_C}
    \begin{align}
        C^{(0011)}_\mathrm{GW} & =\frac{8\left(64+9 \pi ^2\right)}{45}\approx 27.16,\label{eq:GW_single_0011}\\
        C^{(1011)}_\mathrm{GW} & =\frac{64+9 \pi ^2}{360} \approx 42.45\times10^{-2},\label{eq:GW_single_1011}\\
        C^{(0122)}_\mathrm{GW} & =\frac{4096+1225 \pi ^2 }{1433600} \approx 11.29\times10^{-3},\\
         C^{(1122)}_\mathrm{GW} & =\frac{8 \left(4096+1225 \pi ^2\right)}{93002175} \approx 13.92\times10^{-4}.
    \end{align}
\end{subequations}
The coefficient $C^{(0011)}_\mathrm{GW}$ has been calculated in Ref.~\cite{Baryakhtar:2017ngi} by solving the linear Einstein equation using the Green's function. The estimated value is 6.4 in the flat background spacetime and $60$ in the Schwarzschild geometry. By fitting the numerical results, the authors of Ref.~\cite{Siemonsen:2019ebd} obtain a value of $C^{(0011)}_\mathrm{GW}=16.66$. Our result is consistent with these previous calculations. Note that Eq.~\eqref{eq:GW_single} is the radiation of each single mode, without the coannihilation of two vectors from different modes.

\begin{figure}[htbp]
    \centering
    \includegraphics[width=0.48\textwidth]{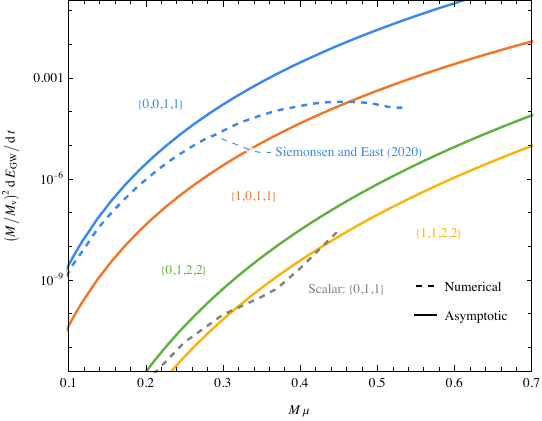}
    \caption{The GW emission fluxes as functions of the mass coupling $M\mu$ for four superradiant modes $^3S_{1,1}^0$ (blue), $^3S_{1,1}^1$ (red), $^3P_{2,2}^0$ (green) and $^3P_{2,2}^1$ (orange). The solid curves are from \eref{eq:GW_single} with Eqs.~\eqref{eq:GW_single_C}, assuming $M\mu\ll1$ and no interference between different modes. The numerical results for the $^3S_{1,1}^0$ mode (blue dashed) \cite{Siemonsen:2019ebd} and scalar $^1P_{1,1}^0$ mode (gray dashed) \cite{Guo:2022mpr} are also shown for comparison.}
    \label{fig:GW_emission}
\end{figure}

In \figref{fig:GW_emission}, we plot the GW emission fluxes in \eref{eq:GW_single}  as functions of $M\mu$ with the coefficients given in Eqs.~\eqref{eq:GW_single_C}. Also shown is the numerical result of the $^3S_{1,1}^0$ mode from Ref.~\cite{Siemonsen:2019ebd}. For  $M\mu\ll1$,  our analytical calculation agrees with the numerical solution. At $M\mu=0.1$, the former exceeds the latter by a factor of 1.61, which increases to 6.09 at $M\mu=0.3$ and to 19.69 at $M\mu=0.4$. For comparison, we plot in Fig.~\ref{fig:GW_emission} the strongest GW emission rate of a scalar superradiant condensate, which is from the $^1P_{1,1}^0$ mode. The curve for the scalar field has a $(M\mu)^{14}$ dependence. It aligns with the curve of the vector $^3P_{2,2}^1$ mode, smaller by more than five orders of magnitude compared to the vector $^3S_{1,1}^0$ mode.

Next, we discuss the GW emission with multiple modes. As explained in Sec.~\ref{subsec:srrate}, with a certian $M\mu$ and $a_*$, the dominant mode is $^3(j-1)_{j,j}^0$ and the subdominant mode is $^3(j-1)_{j,j}^1$, where $j$ is the smallest integer satisfying $j>\omega_{0(j-1)jj}/\Omega_H$. 
If all modes have similar initial mass, later the dominant mode will have the largest number of particles due to its largest superradiant rate among all modes. In this work, we focus on the dominant and subdominant modes, while other modes with a much smaller number of particles are safely ignored.

Defining $\omega_d$ and $\omega_s$ as the frequencies of the dominant and subdominant modes, naively there are four possible frequencies of the emitted gravitons, corresponding to, 
\begin{enumerate}
    \item Annihilation of two dominant vectors, producing a graviton with frequency $\widetilde{\omega}_1 = 2\,\omega_d$;
    \item Annihilation of two subdominant vectors, producing a graviton with frequency $\widetilde{\omega}_2 =2\,\omega_s$;
    \item Annihilation of a dominant vector and a subdominant vector, producing a graviton with frequency $\widetilde{\omega}_3 =\omega_d+\omega_s$;
    \item The transition of a vector from the subdominant mode to the dominant mode, producing a graviton with frequency $\widetilde{\omega}_4 =\omega_s-\omega_d$.
\end{enumerate}
Defining $N_d$ and $N_s$ as the numbers of particles in the dominant and the subdominant modes, one could obtain the strengths of these four processes. The amplitudes of the first two processes are proportional to $N_d^2$ and $N_s^2$, respectively. The corresponding GW emission fluxes are given in Eq.~\eqref{eq:GW_single} above. The last two processes are proportional to $N_d N_s$, with the transition process further suppressed by the graviton phase space.

These processes change the number of vectors in different modes, which are important in the time evolution of the BH-condensate. For $N_d \gg N_s$, the loss of the dominant-mode vectors is mainly via the first process. The loss of the subdominant-mode vectors is mainly via the third process. Using the stress tensor with two modes in the calculation explained in Ref.~\cite{Brito:2014wla}, one could obtain the GW flux from the third process. For the subdominant modes with $j=1$ and $j=2$, the results are the following:
\begin{subequations}\label{eq:GW_sub_int}
\begin{align}
    \dot{E}^{(1011)}_\mathrm{GW} &\approx \frac{2(64+9\pi^2)}{45} \frac{M_\mathrm{v}^{(0011)}M_\mathrm{v}^{(1011)}}{M^2} (M\mu)^{10}, \label{eq:GW_single_1011_new}\\
       \dot{E}^{(1122)}_\mathrm{GW} &\approx \frac{4096+1225 \pi ^2}{2041200} \frac{M_\mathrm{v}^{(0122)}M_\mathrm{v}^{(1122)}}{M^2} (M\mu)^{14}. \label{eq:GW_single_1122_new}
\end{align}
\end{subequations}
In the evolution of the BH-condensate below, we keep only the most important process for each mode. Specifically, we use Eq.~\eqref{eq:GW_single} for dominant modes and Eqs.~\eqref{eq:GW_sub_int} for subdominant modes.

The coexistence of multiple modes has another important consequence. Due to the small difference in the frequencies of the two modes, the dominant and the subdominant modes interfere, producing a unique beat signature in the GW emission flux. This beat signature from BH-scalar-condensate has been studied in detail in Ref.~\cite{Guo:2022mpr}. In this work, we study the similar signature of the BH-vector-condensate systems. We first replace $\mathcal{T}_{\mu\nu}$ in Ref.~\cite{Guo:2022mpr} by the one with a Proca field,
\begin{align}
    \begin{split}
        & \mathcal{T}_{\mu\nu}(\mathcal{A}_{(i)},\mathcal{A}_{(j)})\\
        & =  \frac{1}{2}\Bigg[ \mathcal{F}^{(i)}_{\mu\sigma}\mathcal{F}^{(j)}_{\nu}{}^{\sigma}-\mu^2\mathcal{A}^{(i)}_{\mu}\mathcal{A}^{(j)}_{\nu}\\
        & \hspace{2.5em} + g_{\mu\nu}\left(-\frac{1}{4}\mathcal{F}^{(i)}_{\rho\sigma}\mathcal{F}_{(j)}^{\rho\sigma}+\frac{1}{2}\mu^2\mathcal{A}^{(i)}_{\rho}\mathcal{A}_{(j)}^{\rho}\right) + i \leftrightarrow  j \Bigg],
    \end{split}
\end{align}
where the labels $i,j=d$ or $s$, representing the dominant mode or the subdominant mode respectively. $\mathcal{A}^\mu_{(i)}$ is defined below \eref{eq:A_multi}, and $\mathcal{F}^{(\nu)}_{\mu\sigma}$ is defined by \eref{eq:field_strength_tens} with $A^\mu$ replaced by $\mathcal{A}^\mu_{(i)}$. Following the same steps in Ref.~\cite{Guo:2022mpr}, one obtains the emission flux when these two modes coexist,
\begin{widetext}
    \begin{equation}
    \begin{aligned}
    \frac{dE_\text{GW}}{dt}=&\frac{r^2}{16\pi}\int d\Omega \,\Big\langle\dot{h}_+^2+\dot{h}_\times^2\Big\rangle\\
    =&\frac{1}{8\pi}
    {\sum_{\widetilde{l}}}
    \Bigg\{
    \dfrac{N_{d}^{2}}{{\omega_{d}}^2} \frac{\left|U_{\widetilde{l}\widetilde{m}}^{(\widetilde{\omega}_1)}\right|^{2}}{\widetilde{\omega}_{1}^2}
    +\dfrac{N_{s}^{2}}{{\omega_{s}}^2}\frac{\left|U_{\widetilde{l}\widetilde{m}}^{(\widetilde{\omega}_2)}\right|^{2}}{\widetilde{\omega}_{2}^2} 
    +4\dfrac{N_{d}N_{s}}{\omega_{d}\omega_{s}} \frac{\left|U_{\widetilde{l}\widetilde{m}}^{(\widetilde{\omega}_3)}\right|^{2}}{\widetilde{\omega}_{3}^2}\\
    & +4 \sqrt{\dfrac{N_{d}^3N_{s}}{{\omega_{d}}^3\omega_{s}}} 
    \frac{\left|U_{\widetilde{l}\widetilde{m}}^{(\widetilde{\omega}_1)}\right| \left|U_{\widetilde{l}\widetilde{m}}^{(\widetilde{\omega}_3)}\right|}{\widetilde{\omega}_{1} \widetilde{\omega}_{3}}\cdot \cos\left[\widetilde{\omega}_4\left(t-r\right)
    -\phi_{\widetilde{l}\widetilde{m}}^{(\widetilde{\omega}_3)}+\phi_{\widetilde{l}\widetilde{m}}^{(\widetilde{\omega}_1)}
    \right] \\
    & +2 \dfrac{N_{d}N_{s}}{\omega_{d}\omega_{s}} 
    \frac{\left|U_{\widetilde{l}\widetilde{m}}^{(\widetilde{\omega}_1)}\right| \left|U_{\widetilde{l}\widetilde{m}}^{(\widetilde{\omega}_2)}\right|}{\widetilde{\omega}_{1} \widetilde{\omega}_{2}}\cdot {\cos\left[2\widetilde{\omega}_4\left(t-r\right)
    -\phi_{\widetilde{l}\widetilde{m}}^{(\widetilde{\omega}_2)}+\phi_{\widetilde{l}\widetilde{m}}^{(\widetilde{\omega}_1)}
    \right]} \\
    & +4 \sqrt{\dfrac{N_{d}N_{s}^3}{{\omega_{d}}{\omega_{s}}^3}} 
    \frac{\left|U_{\widetilde{l}\widetilde{m}}^{(\widetilde{\omega}_2)}\right| \left|U_{\widetilde{l}\widetilde{m}}^{(\widetilde{\omega}_3)}\right|}{\widetilde{\omega}_{2} \widetilde{\omega}_{3}}\cdot {\cos\left[\widetilde{\omega}_4\left(t-r\right)
    -\phi_{\widetilde{l}\widetilde{m}}^{(\widetilde{\omega}_2)}+\phi_{\widetilde{l}\widetilde{m}}^{(\widetilde{\omega}_3)}
    \right]} 
    \Bigg\}.
      \end{aligned}
      \label{eq:dE/dt_beat}
    \end{equation}
\end{widetext}
where $\widetilde{l}$ and $\widetilde{m}$ are the orbital and azimuthal numbers for the emitted GW, respectively. Specifically with the dominant mode $^3(j-1)_{j,j}^0$ and subdominant mode $^3(j-1)_{j,j}^1$ under consideration, one has $\widetilde{m}=2j$ and  $\widetilde{l} \geq \widetilde{m}$. The phase angle $\widetilde{\phi}^{(\widetilde{\omega})}_{\widetilde{l} \widetilde{m}}$ is defined as $\arg \left(U^{(\widetilde{\omega})}_{\widetilde{l} \widetilde{m}}\right)$. Finally, the most important $U^{(\widetilde{\omega})}_{\widetilde{l} \widetilde{m}}$ are
\begin{align}
U_{22}^{(\widetilde{\omega}_1)} \approx & \frac{16}{3} \sqrt{\frac{\pi }{5}} (8+3 i \pi )  M^4\mu ^8, \\
U_{22}^{(\widetilde{\omega}_2)} \approx & \frac{2}{3} \sqrt{\frac{\pi }{5}} (8+3 i \pi ) M^4 \mu ^8,  \\
U_{22}^{(\widetilde{\omega}_3)} \approx & \frac{4}{3} \sqrt{\frac{2 \pi }{5}} (8+3 i \pi )   M^4\mu ^8
\end{align}
for $j=1$, and
\begin{align}
U_{44}^{(\widetilde{\omega}_1)} \approx & -\frac{1}{80}\sqrt{\frac{\pi }{7}} (64+35 i \pi )  M^6\mu ^{10} , \\
U_{44}^{(\widetilde{\omega}_2)} \approx & -\frac{16 (64+35 i \pi )}{3645} \sqrt{\frac{\pi }{7}} M^6\mu ^{10}, \\
U_{44}^{(\widetilde{\omega}_3)} \approx & -\frac {1} {135} \sqrt {\frac {\pi} {7}} (64 + 35 i\pi) M^6\mu^{10}
\end{align}
for $j=2$. Here only the leading order terms in $M\mu$ are kept. The contributions with $\widetilde{l}>\widetilde{m}$ are much smaller and can be safely ignored.

In Eq.~\eqref{eq:dE/dt_beat}, the six terms in the curly bracket are the square of contributions from the first three processes as well as their cross terms. Due to the small difference of $\omega_d$ and $\omega_s$, the cross terms result in the beat signals with frequencies $\widetilde{\omega}_4$ and $2\,\widetilde{\omega}_4$. The fourth process does not exist in Eq.~\eqref{eq:dE/dt_beat} because the dominant and subdominant modes have the same value of $m$. In the case $N_d\gg N_s$, the first and the fourth terms in the curly brackets are the most important with other terms suppressed by more powers of $(N_s/N_d)^{1/2}$. In this limit, the GW flux is monochromatic with a beat frequency $\widetilde{\omega}_4$. As a cross-check, one could set $N_d$ or $N_s$ to be zero, then this equation reduces to \eref{eq:GW_single} with the coefficients in Eqs.~\eqref{eq:GW_single_C}.

\subsection{Evolution equations}
\label{sec:sub_Evo_eqs}

In this part, we derive the evolution equation of the BH-condensate system, with the superradiant rate and the GW emission fluxes obtained in previous subsections.

The superradiance results in the energy flux and the  angular momentum flux through the BH horizon, which are given by
\begin{subequations}
\begin{align}
    \dot{E}_{\mathrm{H}} & = \sum_{nljm} \dot{E}_{\mathrm{H}}^{(nljm)} \equiv \sum_{nljm} 2 M^{(nljm)}_\mathrm{v} \Gamma_{nljm},\label{eq:dEH_dt}\\
    \dot{J}_{\mathrm{H}} & = \sum_{nljm} \dot{J}_{\mathrm{H}}^{(nljm)} \equiv \sum_{nljm} \frac{m}{\omega_{nljm}} \dot{E}_{\mathrm{H}}^{(nljm)},\label{eq:dJH_dt}
\end{align}
\end{subequations}
where $M^{(nljm)}_\mathrm{v}$ is the total mass of the $^3l_{j,m}^n$ mode. The factor $m/\omega_{nljm}$ arises because every vector particle has the energy $\omega_{nljm}$ and the angular momentum $m$ along the $z$ direction observed at infinity \cite{Bekenstein:1973mi}. 

The direction of the flux can be inferred from the sign of the superradiance rate. The flux flows out of the horizon when $\Gamma_{nljm}>0$ and into the horizon when $\Gamma_{nljm}<0$. According to \eref{eq:SR_rate_Baumann}, the superradiance condition $\Gamma_{nljm}>0$ is equivalent to $\omega_{nljm}<m\Omega_\mathrm{H}$. This allows us to derive the critical BH spin for each mode,
\begin{align}\label{eq:a_critc}
    a^{(nljm)}_\mathrm{*c}=\frac{4mM\omega_{nljm}}{m^2+\left( 2M\omega_{nljm} \right) ^2}.
\end{align}
Superradiance of the $^3l_{j,m}^n$ mode exists only when the BH spin is above this value. With the presence of the $^3l_{j,m}^n$ mode, there is an attractor of the BH spin. If BH spin is below this value, the angular momentum flows into the horizon. As a result, the BH spin increases while the $^3l_{j,m}^n$ mode shrinks. Conversely, if the BH spin is above this critical value, the angular momentum flows from the BH to the condensate, reducing the BH spin while increasing the size of the $^3l_{j,m}^n$ mode. In the limit of $M\mu\ll1$, \eref{eq:a_critc} has the expanded form
\begin{align}
    a^{(nljm)}_\mathrm{*c}=\frac{4}{m}(M\mu)-\frac{2 \left(m^2+8 \bar{n}^2\right)}{m^3 \bar{n}^2}(M\mu)^3 + \mathcal{O}[(M\mu)^4].
    \label{eq:a_critc_app}
\end{align}

Meanwhile, the $^3l_{j,m}^n$ mode loses its energy and angular momentum continuously via GW emission. The two loss-rates are related by
\begin{align}
    \dot{J}_{\mathrm{GW}}^{(nljm)}=\frac{\widetilde{m}}{\widetilde{\omega}}\dot{E}_{\mathrm{GW}}^{(nljm)},
\end{align}
where $\widetilde{\omega}$ and $\widetilde{m}$ are the GW energy and angular momentum, respectively. The energy loss rates of the two dominant modes are given in Eq.~\eqref{eq:GW_single}. The energy loss of the subdominant modes is mainly from the interference with the dominant ones, which are provided in Eq.~\eqref{eq:GW_sub_int}.

Finally, the evolution equations of the BH-condensate system are
\begin{subequations}\label{eq:Evoluton_ODE}
\begin{align}
    \dot{M}&=-\dot{E}_{\mathrm{H}},\\
    \dot{J}&=-\dot{J}_{\mathrm{H}},\\
    \dot{M}_{\mathrm{v}}^{(nljm)} & = \dot{E}_{\mathrm{H}}^{(nljm)} -\dot{E}_{\mathrm{GW}}^{(nljm)},\label{eq:Evoluton_E}\\
    \dot{J}_{\mathrm{v}}^{(nljm)} & =\dot{J}_{\mathrm{H}}^{(nljm)} - \dot{J}_{\mathrm{GW}}^{(nljm)},
\end{align}
\end{subequations}
where $J$ and $M$ are the angular momentum and mass of the BH, respectively, and $J^{(nljm)}_\mathrm{v}$ is the angular momentum of the $^3 l_{j,m}^n$ mode. 

In obtaining these evolution equations, we have made the following assumptions:
\begin{itemize}
    \item 
    The superradiance rate in Eq.~\eqref{eq:SR_rate_Baumann} and the GW emission fluxes in Eqs.~\eqref{eq:GW_single} and \eqref{eq:GW_sub_int} require $M\mu\ll 1$.
    \item 
    The backreaction of the condensate on the metric is ignored due to the low energy density of the vector condensate. This simplification is qualified since the Bohr radius of the condensate $r_\mathrm{B}$ is much larger than the BH outer horizon $r_+$ in the $M\mu\ll 1$ limit.
    \item 
    The self-interaction of the vector field is ignored, also owing to its low energy density.
    \item 
    The quasi-adiabatic approximation is employed, which assumes that the dynamical timescale of the BH is much shorter than the timescales of both the superradiant instability and the GW emission \cite{Brito:2014wla,Brito:2017zvb}.
    \item 
     The BH-condensate system is assumed to be isolated. The effects of accretion or other close-by objects are beyond the scope of this work.
\end{itemize}

\section{Evolution with a single dominant mode}
\label{sec:Single-mode}

The dominant modes have $l=j-1$. Then the superradiance rate in Eq.~\eqref{eq:SR_rate_Baumann} is proportional to $(M\mu)^{4j+3}$, while the GW emission rate in Eq.~\eqref{eq:GW_single} is proportional to $(M\mu)^{4j+6}$. In the limit of $M\mu\ll 1$, the superradiance process is much faster than the energy dissipation. This hierarchy results in two well-separated phases in the evolution of the dominant modes. In the first phase, the condensate extracts energy and angular momentum from the host BH. As a result, the BH spin drops until reaching the critical value $a_{*\text{c}}$ and the superradiance terminates. The evolution of this phase is governed by superradiance while GW emission can be safely ignored. We refer to this phase as the {\it spin-down phase}. Then in the second phase, the condensate slowly loses mass and angular momentum via GW emission. The BH mass and spin are unchanged, which is an attractor mathematically. We refer to this phase as the {\it attractor phase}.

This separation qualifies an approximate analysis of the BH-condensate evolution. In this section, we take the $^3S_{1,1}^0$ mode as an example, giving analytic approximations of the BH mass and spin, the maximum condensate mass, and the durations of the two phases. These approximations are then compared to the numerical results obtained by solving Eqs.~\eqref{eq:Evoluton_ODE} directly. we further discuss the effects of the initial BH spin, the initial condensate mass, and the initial mass coupling on the BH-condensate evolution.

\subsection{Analytic approximation}\label{subsec:single_analytic}

In the spin-down phase, the GW emission can be safely ignored. The initial BH mass and spin are $M_0$ and $a_{*0}$, respectively. The initial mass of the condensate is $M_\mathrm{v0}^{(0011)}$. Then the condensate extracts energy and angular momentum from the host BH. The spin-down phase ends at time $t_1$, when the BH spin decreases to $a^{(0011)}_{*\text{c}}$ in Eq.~\eqref{eq:a_critc} and the BH mass drops to $M_1$. At the same time, the condensate reaches its maximum mass $M^{(0011)}_{\text{v,max}}$. The conservations of energy and angular momentum are
\begin{subequations}\label{eq:e_am_conservation}
    \begin{align}
    M_0 & = M_1 + (M_\mathrm{v,max}^{(0011)}-M_\mathrm{v0}^{(0011)}),\label{eq:M1_single}\\
    a_{*0} M_0^2 & =a^{(0011)}_{*\text{c}} M_1 ^2 + (M_\mathrm{v,max}^{(0011)}-M_\mathrm{v0}^{(0011)})/\omega_{0011},
    \end{align}
    \end{subequations}
where $\omega_{0011}$ is calculated in Eq.~\eqref{eq:omega} with $\alpha$ replaced by $M_1\mu$. By fitting the numerical results from solving the conservation equations, $M_\mathrm{v,max}^{(0011)}$ can be expressed as a power series of $M_0\mu$
\begin{align}\label{eq:Mv_0011_max}
    \begin{split}
        \frac{M_\mathrm{v,max}^{(0011)}}{M_0} \approx & \frac{M_\mathrm{v0}^{(0011)}}{M_0} + M_0\mu \Big[ a_{*0}-4 M_0\mu + 7.82 a_{*0} (M_0\mu)^2 \\
        & \hspace{1cm}+(-14.47 + 5.73 a_{*0}^2)(M_0\mu)^3\Big].
    \end{split}
\end{align}
Then the BH mass  $M_1$ at the end of this phase can be calculated with Eq.~\eqref{eq:M1_single}. The obtained $M_1$ depends only on $a_{*0}$ and $M_0\mu$.

The condensate mass at the end of the spin-down phase can also be calculated by solving Eq.~\eqref{eq:Evoluton_E} with $\dot{E}^{(nljm)}_{\text{GW}}$ set to zero. The difficulty lies in the fact that $\dot{E}_\text{H}^{(0011)}$ depends on $a_*$ via the superradiance rate $\Gamma_{0011}$. From Fig.~\ref{fig:SR_rate_diff_aStar}, $\Gamma_{0011}$ depends only weakly on $a_*$ except in the close neighborhood of $a_{*\text{c}}$. Consequently, one could replace $a_*$ in $\Gamma_{0011}$ by $a_{*0}$ as a good approximation. The equation is then trivial to solve, resulting in exponential growth of the condensate mass,
\begin{align}\label{eq:Mv_max_vs_t}
    M_\mathrm{v,max}^{(0011)} \approx M_\mathrm{v0}^{(0011)}e^{2\Gamma_{0011}t_1}.
\end{align}
Then the duration of this phase can be calculated as
\begin{align}\label{eq:t_1}
    t_1 \approx \tau_\mathrm{sr}^{(0011)}\log\frac{M_\mathrm{v,max}^{(0011)}}{M_\mathrm{v0}^{(0011)}},
\end{align}
where $\tau_\text{sr}^{(0011)}$ is the superradiant timescale,
\begin{align}\label{eq:tau_sr_0011}
    \begin{split}
        \tau_\mathrm{sr}^{(0011)} & = \frac{1}{2\Gamma_{0011}} ,\\
        & \approx 2.05\ \mathrm{min} \left(\frac{M_0}{10M_\odot}\right) \left(\frac{0.1}{M_0\mu}\right)^{7} \left(\frac{1}{a_{*0}}\right).
    \end{split}
\end{align}

The attractor phase starts when the condensate reaches its maximum mass. At this moment, the BH spin is almost $a_{*\text{c}}^{(0011)}$ and superradiance is negligible. The BH mass and spin are then unchanged in this phase. Meanwhile, the condensate dissipates energy and angular momentum via GW emission. The condensate mass in this phase can be solved from Eq.~\eqref{eq:Evoluton_E} with $\dot{E}_\text{H}^{(nljm)}$ set to zero. Inserting \eref{eq:GW_single}, one arrives at
\begin{align}\label{eq:single-decay}
\begin{split}
M_\mathrm{v}^{(0011)}(t) & =
\frac{M_\mathrm{v,max}^{(0011)}}{1+(t-t_1)/\tau_\mathrm{GW}^{(0011)}},
\end{split}
\end{align}
with
\begin{align}\label{eq:tau_GW_0011}
\begin{split}
\tau_\mathrm{GW}^{(0011)} &= \frac{M_1^2}{M_\mathrm{v,max}^{(0011)} C_\mathrm{GW}^{(0011)}(M_1\mu)^{10}}\\
&\approx 
2.10\ \mathrm{days} \left(\frac{M_0}{10M_\odot}\right)\left(\frac{0.1}{M_0\mu}\right)^{10}\left(\frac{0.1}{{M_\mathrm{v,max}^{(0011)}}/M_0}\right).
\end{split}
\end{align}
When $t\gg \tau_\text{GW}^{(0011)}$, since $ \tau_\text{GW}^{(0011)} \gg \tau_\text{sr}^{(0011)}\sim t_1$, the expression reduces to
\begin{align}\label{eq:Mv_large_t}
M_\mathrm{v}^{(0011)}(t) \approx \frac{M_1^2}{ C_\mathrm{GW}^{(0011)}(M_1\mu)^{10}} \frac{1}{t},
\end{align}
which is inversely proportional to $t$ and depends only on $M_1$ and $M_1\mu$.

\subsection{Numerical Calculation}

In this part, we numerically solve Eqs.~\eqref{eq:Evoluton_ODE} for the evolution of the BH-vector-condensate system. The effects of the initial BH spin, the initial condensate mass as well as the initial mass coupling are studied in detail. The obtained analytic approximations are compared to the numerical results to estimate the errors of the former.

\subsubsection{The effect of initial BH spin}

\begin{figure*}
    \centering
    \includegraphics[width=\textwidth]{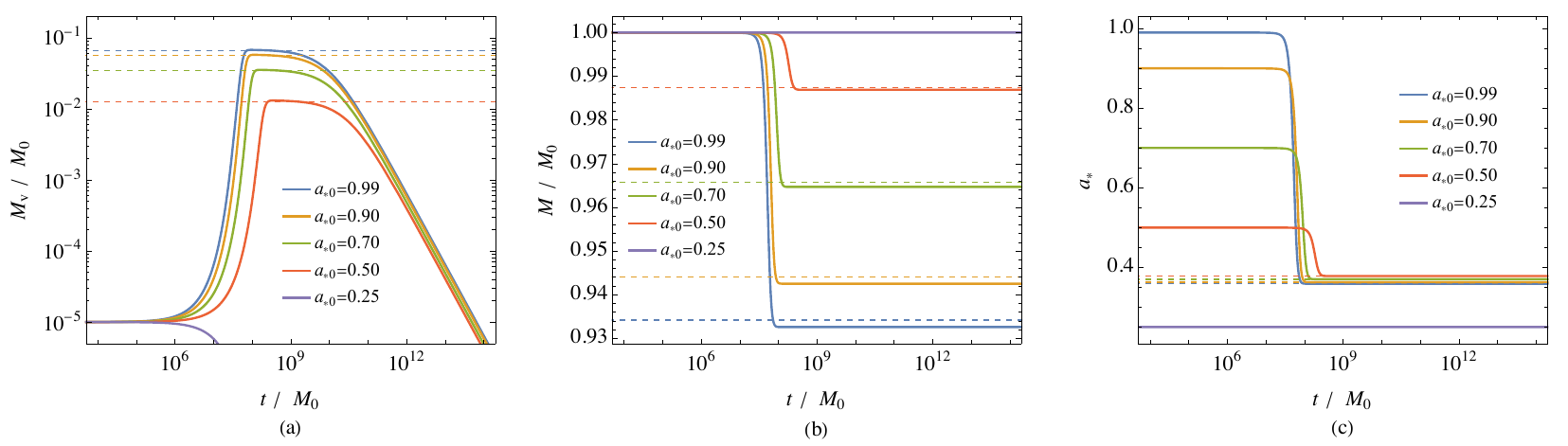}
    \caption{
    The evolutions of the condensate mass (panel a), BH mass (panel b), and BH spin (panel c) as functions of time with different BH initial spins $a_{*0}$, where only the $^3S_{1,1}^0$ mode is included in the evolution. The initial mass coupling is $M_0\mu=0.1$ and the initial mass of the condensate is $10^{-5}M_0$. The solid curves are the numerical solutions from solving the evolution equations in Eqs.~\eqref{eq:Evoluton_ODE}. The dashed horizontal lines in the three panels are from the analytic approximations in Eqs.~\eqref{eq:Mv_0011_max}, \eqref{eq:M1_single} and \eqref{eq:a_critc_app}, respectively. For completeness, the evolution with initial BH spin  $a_{*0}=0.25$ is also shown, which does not satisfy the superradiance condition $\omega_{0011}<\Omega_\mathrm{H}$.}
    \label{fig:1011_ini_as}
\end{figure*}
\begin{table*}
    \renewcommand\arraystretch{1.2}
\begin{centering}
    \begin{tabular}{c|ccccc|ccccc}
        \hline 
        \multirow{2}{*}{$a_{*0}$} & \multicolumn{5}{c|}{Estimates} & \multicolumn{5}{c}{Numerical results}\tabularnewline
         & $M_{\mathrm{v},\max}^{(0011)}/M_{0}$ & $M_1/M_{0}$ & $a_{*\mathrm{c}}$ & $t_1/M_{0}$ & $\tau_{\mathrm{GW}}^{(0011)}/M_{0}$ & $M_{\mathrm{v},\max}^{(0011)}/M_{0}$ & $M_1/M_{0}$ & $a_{*\mathrm{c}}$ & $t_1/M_{0}$ & $\tau_{\mathrm{GW}}^{(0011)}/M_{0}$\tabularnewline
        \hline 
        0.99 & $6.59\times10^{-2}$ & 0.934 & 0.359 & $4.79\times10^{7}$ & $9.64\times10^{9}$ & $6.71\times10^{-2}$ & 0.933 & 0.359 & $9.51\times10^{7}$ & $9.69\times10^{9}$\tabularnewline
        0.90 & $5.61\times10^{-2}$ & 0.944 & 0.362 & $6.19\times10^{7}$ & $1.04\times10^{10}$ & $5.72\times10^{-2}$ & 0.943 & 0.363 & $1.12\times10^{8}$ & $1.04\times10^{10}$\tabularnewline
        0.70 & $3.43\times10^{-2}$ & 0.966 & 0.370 & $8.88\times10^{7}$ & $1.42\times10^{10}$ & $3.52\times10^{-2}$ & 0.965 & 0.370 & $1.60\times10^{8}$ & $1.41\times10^{10}$\tabularnewline
        0.50 & $1.26\times10^{-2}$ & 0.987 & 0.378 & $1.81\times10^{8}$ & $3.23\times10^{10}$ & $1.30\times10^{-2}$ & 0.987 & 0.378 & $3.56\times10^{8}$ & $3.18\times10^{10}$\tabularnewline
        \hline 
    \end{tabular} 
\par\end{centering}
    \caption{
Comparison of the analytic approximations to the numerical solutions for the important quantities with different initial BH spins $a_{*0}$, where the only the $^3S_{1,1}^0$ mode is included in the evolution. The initial mass coupling is $M_0\mu=0.1$ and the initial mass of the condensate is $10^{-5}M_0$. The numerical solutions are from solving Eqs.~\eqref{eq:Evoluton_ODE}. The estimates of the maximum condensate mass $M_\text{v,max}^{(0011)}$, the final BH mass $M_1$, the final BH spin $a_{*\text{c}}$, the duration of the spin-down phase $t_1$ and the GW emission time scale $\tau_{\mathrm{GW}}^{(0011)}$ are calculated with Eqs.~\eqref{eq:Mv_0011_max}, \eqref{eq:M1_single}, \eqref{eq:a_critc_app}, \eqref{eq:t_1} and \eqref{eq:tau_GW_0011}, respectively.}
    \label{table:estimate_spin}
\end{table*}

Fig.~\ref{fig:1011_ini_as} shows the condensate mass, the BH mass and spin as functions of time. The fixed parameters are $M_0\mu=0.1$ and $M_\text{v0}=10^{-5}M_0$. Four different initial BH spins $a_{*0}=0.99$, $0.9$, $0.7$, $0.5$ are then compared. As expected, each curve can be separated into two phases. In the spin-down phase, the condensate mass presents an exponential behavior. A larger value of $a_{*0}$ leads to a larger maximum condensate mass, consistent with Eq.~\eqref{eq:Mv_0011_max}. The BH loses its mass and spin in this phase.

The attractor phase starts when the condensate reaches the maximum mass. In this phase, the condensate loses energy via GW emission. The BH mass and spin are apparently unchanged. Especially, the BH spin equals to $a_{*\text{c}}^{(0011)}$ determined by Eq.~\eqref{eq:a_critc}, with $M$ replaced by $M_1$ in each case.

The analytic approximations are compared to the numerical values in Table \ref{table:estimate_spin}. The errors are all less than 10\% except for $t_1$. Nonetheless, Eq.~\eqref{eq:t_1} still gives the correct order of magnitude of $t_1$.

For completeness, we also study the evolution with $a_{*0}=0.25$, which is below the critical value $a^{(0011)}_{*\text{c}}$. As expected, the condensate is absorbed by the BH. Since the initial condensate mass is only $10^{-5}M_0$, the changes of BH mass and spin are unobservable in Fig.~\ref {fig:1011_ini_as}.

\subsubsection{The effect of initial condensate mass values}\label{subsec:Mv0}

\begin{figure*}
    \centering
    \includegraphics[width=\textwidth]{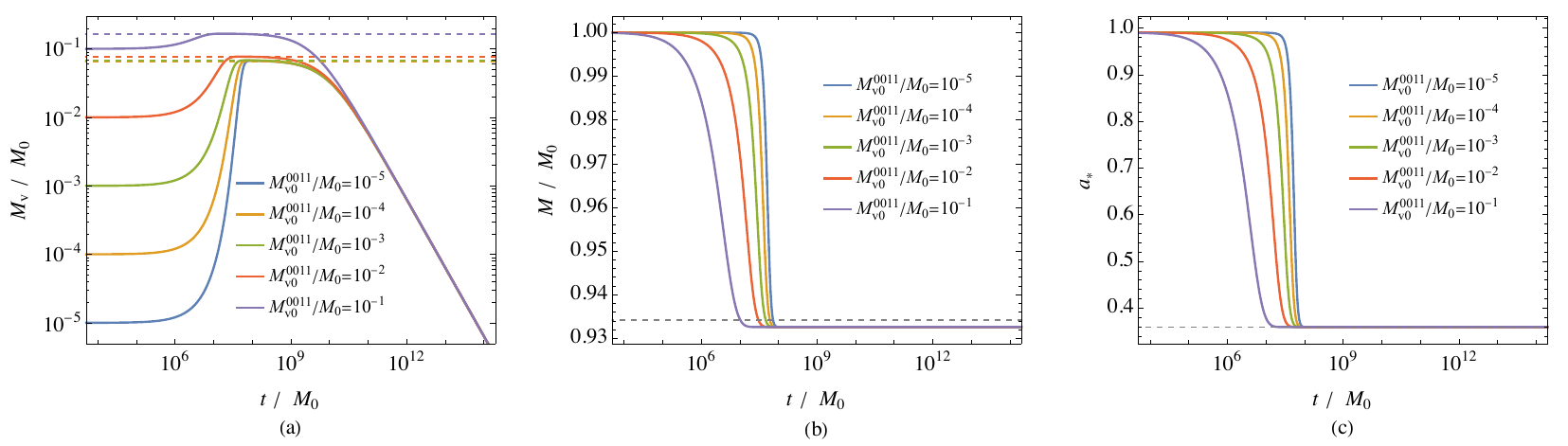}
    \caption{The evolutions of the condensate mass (panel a), BH mass (panel b), and BH spin (panel c) as functions of time with different initial condensate masses $M_\text{v0}^{(0011)}$, where only the $^3S_{1,1}^0$ mode is included in the evolution. The initial mass coupling is $M_0\mu=0.1$ and the initial BH spin is $a_{*0}=0.99$. The solid curves are the numerical solutions from solving the evolution equations in Eqs.~\eqref{eq:Evoluton_ODE}. The dashed horizontal lines in the three panels are from the analytic approximations in Eqs.~\eqref{eq:Mv_0011_max}, \eqref{eq:M1_single} and \eqref{eq:a_critc_app}, respectively.}
    \label{fig:image/1011_ini_Mv0}
\end{figure*}

\begin{table*}
    \renewcommand\arraystretch{1.2}
\begin{centering}
    \begin{tabular}{c|ccccc|ccccc}
        \hline 
        \multirow{2}{*}{$\dfrac{M_{\mathrm{v}0}^{(0011)}}{M_{0}}$} & \multicolumn{5}{c|}{Estimates} & \multicolumn{5}{c}{Numerical results}\tabularnewline
         & $M_{\mathrm{v},\max}^{(0011)}/M_{0}$ & $M_1/M_{0}$ & $a_{*\mathrm{c}}$ & $t_1/M_{0}$ & $\tau_{\mathrm{GW}}^{(0011)}/M_{0}$ & $M_{\mathrm{v},\max}^{(0011)}/M_{0}$ & $M_1/M_{0}$ & $a_{*\mathrm{c}}$ & $t_1/M_{0}$ & $\tau_{\mathrm{GW}}^{(0011)}/M_{0}$\tabularnewline
        \hline 
        $10^{-5}$ & $6.59\times10^{-2}$ & 0.934 & 0.359 & $4.79\times10^{7}$ & $9.64\times10^{9}$ & $6.71\times10^{-2}$ & 0.933 & 0.359 & $9.51\times10^{7}$ & $9.69\times10^{9}$\tabularnewline
        $10^{-4}$ & $6.60\times10^{-2}$ & 0.934 & 0.359 & $3.53\times10^{7}$ & $9.62\times10^{9}$ & $6.72\times10^{-2}$ & 0.933 & 0.359 & $8.24\times10^{7}$ & $9.66\times10^{9}$\tabularnewline
        $10^{-3}$ & $6.69\times10^{-2}$ & 0.934 & 0.359 & $2.29\times10^{7}$ & $9.49\times10^{9}$ & $6.81\times10^{-2}$ & 0.933 & 0.359 & $6.86\times10^{7}$ & $9.52\times10^{9}$\tabularnewline
        $10^{-2}$ & $7.59\times10^{-2}$ & 0.934 & 0.359 & $1.10\times10^{7}$ & $8.37\times10^{9}$ & $7.71\times10^{-2}$ & 0.933 & 0.359 & $4.88\times10^{7}$ & $8.40\times10^{9}$\tabularnewline
        $10^{-1}$ & $1.66\times10^{-2}$ & 0.934 & 0.359 & $2.75\times10^{6}$ & $3.83\times10^{9}$ & $1.67\times10^{-1}$ & 0.933 & 0.359 & $1.69\times10^{7}$ & $3.88\times10^{9}$\tabularnewline
        \hline 
    \end{tabular}
\par\end{centering}
\caption{Comparison of the analytic approximations to the numerical solutions for the important quantities with different initial condensate mass $M_\text{v0}^{(0011)}$, where only the $^3S_{1,1}^0$ mode is included in the evolution. The initial mass coupling is $M_0\mu=0.1$ and the initial BH spin is $a_{*0}=0.99$. The numerical solutions are from solving Eqs.~\eqref{eq:Evoluton_ODE}. The estimates of the maximum condensate mass $M_\text{v,max}^{(0011)}$, the final BH mass $M_1$, the final BH spin $a_{*\text{c}}$, the duration of the spin-down phase $t_1$ and the GW emission time scale $\tau_{\mathrm{GW}}^{(0011)}$ are calculated with Eqs.~\eqref{eq:Mv_0011_max}, \eqref{eq:M1_single}, \eqref{eq:a_critc_app}, \eqref{eq:t_1} and \eqref{eq:tau_GW_0011}, respectively.}
\label{table:estimate_vector_mass}
\end{table*}

Figure~\ref{fig:image/1011_ini_Mv0} shows the condensate mass, the BH mass and spin as functions of time. The fixed parameters are $a_{*0}=0.99$ and $M_0\mu=0.1$. Five different initial condensate masses $M_\mathrm{v0}^{(0011)}/M_0=10^{-1},10^{-2}, 10^{-3}, 10^{-4}, 10^{-5}$ are then compared. The comparison of the analytic approximations and the numerical results are further presented in Table~\ref{table:estimate_vector_mass}.

As expected, when the initial condensate mass is tuned up, the extraction of energy and spin from the BH is getting faster in the spin-down phase, resulting in a smaller value of $\tau_\text{sr}$. For all cases, the maximum condensate masses agree very well with the analytic approximation in Eq.~\eqref{eq:Mv_0011_max}.

In the attractor phase, the BH mass and spin stay unchanged, with values predicted by the analytic approximations. In particular, the BH mass and spin are independent of the initial condensate mass.  In this phase, the condensate mass drops monotonically due to GW emission. When $t$ is larger than $\sim10^{11} M_0$, the five curves in Fig.~\ref{fig:image/1011_ini_Mv0}(a) merge, which is anticipated in Eq.~\eqref{eq:Mv_large_t}.

Analytic approximations show that $M_1$ and $M_\text{v}$ are independent of $M_\text{v0}$ in the later part of the emission phase. It is confirmed by looking at the first two panels in Fig.~\ref{fig:image/1011_ini_Mv0} with $t\gtrsim 10^{11}M_0$. This finding is important because it shows that the difference in $M_\text{v0}$ is completely eliminated by GW emission. This observation is crucial in our study of multi-mode evolution later.

\subsubsection{The effect of initial mass couplings}

\begin{figure*}
    \centering
    \includegraphics[width=\textwidth]{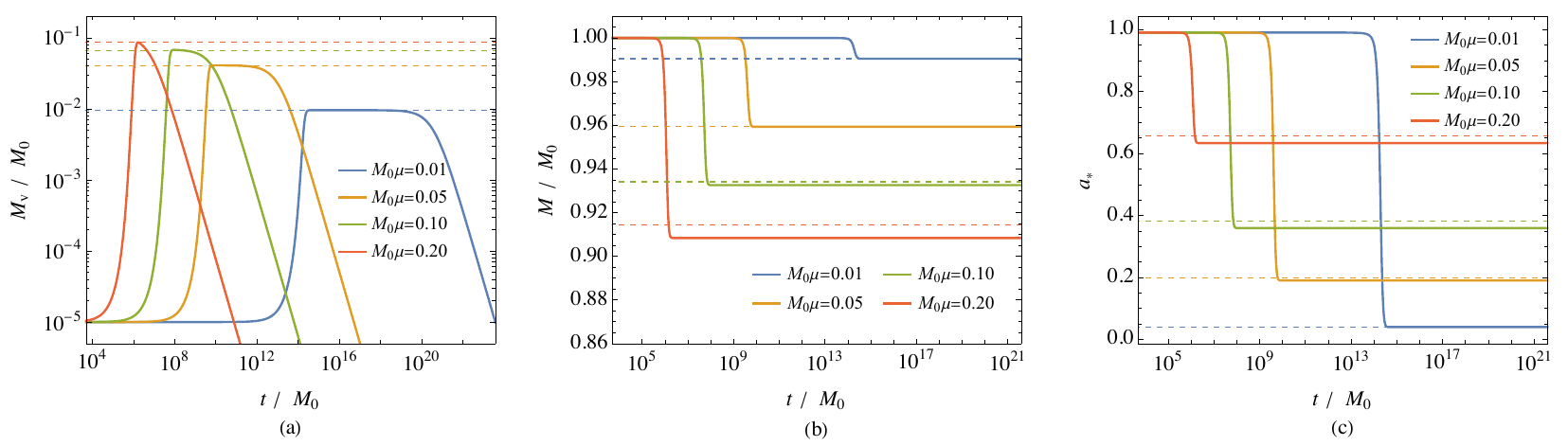}
    \caption{
    The evolutions of the condensate mass (panel a), BH mass (panel b), and BH spin (panel c) as functions of time with different initial mass couplings $M_0\mu$, where only the $^3S_{1,1}^0$ mode is included in the evolution. The initial BH spin is $a_{*0}=0.99$ and the initial condensate mass is $M_\mathrm{v0}^{(0011)}=10^{-5}M_0$. The solid curves are the numerical solutions from solving the evolution equations in Eqs.~\eqref{eq:Evoluton_ODE}. The dashed horizontal lines in the three panels are from the analytic approximations in Eqs.~\eqref{eq:Mv_0011_max}, \eqref{eq:M1_single} and \eqref{eq:a_critc_app}, respectively.}
    \label{fig:1011_ini_alpha0}
\end{figure*}
\begin{table*}
    \renewcommand\arraystretch{1.2}
\begin{centering}
    \begin{tabular}{c|ccccc|ccccc}
        \hline 
        \multirow{2}{*}{$M_{0}\mu$} & \multicolumn{5}{c|}{Estimates} & \multicolumn{5}{c}{Numerical results}\tabularnewline
         & $M_{\mathrm{v},\max}^{(0011)}/M_{0}$ & $M_1/M_{0}$ & $a_{*\mathrm{c}}$ & $t_1/M_{0}$ & $\tau_{\mathrm{GW}}^{(0011)}/M_{0}$ & $M_{\mathrm{v},\max}^{(0011)}/M_{0}$ & $M_1/M_{0}$ & $a_{*\mathrm{c}}$ & $t_1/M_{0}$ & $\tau_{\mathrm{GW}}^{(0011)}/M_{0}$\tabularnewline
        \hline 
        0.01 & $9.52\times10^{-3}$ & 0.990 & 0.0396 & $1.86\times10^{14}$ & $4.17\times10^{20}$ & $9.52\times10^{-3}$ & 0.990 & 0.0396 & $6.40\times10^{14}$ & $4.17\times10^{20}$\tabularnewline
        0.05 & $4.04\times10^{-2}$ & 0.960 & 0.190 & $3.85\times10^{9}$ & $1.30\times10^{13}$ & $4.07\times10^{-2}$ & 0.959 & 0.190 & $8.99\times10^{9}$ & $1.29\times10^{13}$\tabularnewline
        0.10 & $6.59\times10^{-2}$ & 0.934 & 0.359 & $4.79\times10^{7}$ & $9.64\times10^{9}$ & $6.71\times10^{-2}$ & 0.933 & 0.359 & $9.51\times10^{7}$ & $9.69\times10^{9}$\tabularnewline
        0.20 & $8.58\times10^{-2}$ & 0.914 & 0.621 & $1.03\times10^{5}$ & $8.59\times10^{6}$ & $8.56\times10^{-2}$ & 0.914 & 0.633 & $1.69\times10^{6}$ & $1.09\times10^{7}$\tabularnewline
        \hline 
    \end{tabular}
\par\end{centering}
\caption{
Comparison of the analytic approximations to the numerical solutions for the important quantities with different initial mass coupling $M_0\mu$, where only the $^3S_{1,1}^0$ mode is included in the evolution. The initial BH spin is $a_{*0}=0.99$ and the initial mass of the condensate is $10^{-5}M_0$. The numerical solutions are from solving Eqs.~\eqref{eq:Evoluton_ODE}. The estimates of the maximum condensate mass $M_\text{v,max}^{(0011)}$, the final BH mass $M_1$, the final BH spin $a_{*\text{c}}$, the duration of the spin-down phase $t_1$ and the GW emission time scale $\tau_{\mathrm{GW}}^{(0011)}$ are calculated with Eqs.~\eqref{eq:Mv_0011_max}, \eqref{eq:M1_single}, \eqref{eq:a_critc_app}, \eqref{eq:t_1} and \eqref{eq:tau_GW_0011}, respectively.}
\label{table:estimate_mass_coupling}
\end{table*}

Figure~\ref{fig:1011_ini_alpha0} shows the condensate mass, the BH mass and spin as functions of time. Fixed parameters are $a_{*0}=0.99$ and $M_\text{v0}/M_0=10^{-5}$. Five different initial mass couplings $M_0\mu=0.01, 0.05, 0.1, 0.3, 0.4$ are compared. The comparison of the analytic approximations and the numerical results are further presented in Table~\ref{table:estimate_mass_coupling}.

The superradiance rate increases as $(M\mu)^7$. The GW emission rate increases even faster, as $(M\mu)^{10}$. Thus, both $\tau_\text{sr}$ and $\tau_\text{GW}$ decline when $M_0\mu$ gets larger. Although analytic approximations are in good agreement with the numerical results, the errors are larger for larger values of $M_0\mu$. This is because the analytic approximations are Taylor series of $M\mu$. One could keep more terms in the series for better agreement.

Since the emission rate changes more rapidly with $M_0\mu$, its influence in the spin-down phase gets more important. This can be seen in Fig.~\ref{fig:1011_ini_alpha0}(a) when the plateau shape of the curve gets narrower and eventually turns into a peak at $M_0\mu \approx 0.2$. To obtain accurate results for $M_\text{v,max}$, $\tau_\text{sr}$ and $\tau_\text{GW}$ with $M_0\mu\gtrsim 0.2$, it is necessary to include the GW emission in the spin-down phase, which is ignored in the analytic approximations obtained above.

\section{Multi-mode evolution} \label{sec:Multi-mode}

In the realistic evolution of BH-condensate systems, seeds of more than one mode coexist, leading to a multi-mode evolution of the system. In this section, we examine the scenario with two dominant modes ($^3S_{1,1}^0$ and $^3P_{2,2}^0$) and two subdominant modes ($^3S_{1,1}^1$ and $^3P_{2,2}^1$). 

For the $j=1$ modes, the superradiance rate in Eq.~\eqref{eq:SR_rate_Baumann} is proportional to $(M\mu)^7$, while the GW emission rate is proportional to $(M\mu)^{10}$.   For the $j=2$ modes, the superradiance rate is $\sim (M\mu)^{11}$, and the emission rate is $\sim (M\mu)^{14}$. In the $M\mu \ll 1$ limit, the $j=2$ modes become crucial at the time when the $j=1$ modes are almost drained by GW emission. As a result, the evolution of these four modes can be separated into two stages. In the first stage, only the $j=1$ modes are important while the $j=2$ modes can be ignored. Then in the second stage, the $j=1$ modes are almost depleted and one only needs to consider the $j=2$ modes.  Such hierarchy also leads to different phases in each stage similar to the single-mode evolution. In fact, we will see that the effect of the subdominant mode is so small that the evolutions of the dominant mode and the BH are almost the same as the single-mode case detailed in Sec.~\ref{sec:Single-mode}.

In this section, we first study the $j=1$ stage in detail, focusing on the effect of the subdominant mode. Analytic approximations of important quantities are obtained. Then we move on to the $j=2$ stage, concentrating on its similarities and the differences from the $j=1$ stage.  Finally, we numerically solve the evolution equations and compare the results with the analytic approximations. We assume the BH spin is larger than $a_{*\text{c}}^{(1011)}$ so that the $j=1$ modes could grow via superradiance. Otherwise one should start from the minimal integer $j$ which satisfies $j>\omega_{0(j-1)jj}/\Omega_H$. Nonetheless, the superradiance with small BH initial spin $a_{*0}$ is not of much interest phenomenologically.

\subsection{The $j=1$ stage}
\label{sec:m=1_stage}

In this stage, we consider the dominant mode $^3S_{1,1}^0$ and the subdominant mode $^3S_{1,1}^1$. There are two differences between these two modes. Firstly, from \figref{fig:SR_rate_diff_n}, the superradiance rate of the $^3S_{1,1}^0$ mode is approximately an order of magnitude greater than that of the $^3S_{1,1}^1$ mode. A more careful study with Eq.~\eqref{eq:SR_rate_Baumann} indicates $\Gamma_{0011}/\Gamma_{1011}\approx 8$. Secondly, the critical BH spin $a_{*\text{c}}$ of the $^3S_{1,1}^0$ mode is slightly smaller than that of the $^3S_{1,1}^1$ mode, indicated by Eq.~\eqref{eq:a_critc_app}.

With the subdominant mode, the evolution of this stage can be separated into three phases. In the first phase, the BH spin drops from the initial value at $t=0$ to $a_{*\text{c}}^{(1011)}$ at $t_1$. During this time, both modes grow almost exponentially,
\begin{align}\label{eq:two_modes_exp}
M_\mathrm{v}^{(i)}(t_1) \approx M_\mathrm{v0}^{(i)}e^{2\Gamma_{i} t_1},
\end{align}
where $i=0011 (1011)$ for the dominant (subdominant) mode. It is useful to consider the ratio of the two masses at $t_1$,
\begin{align} \label{eq:Max_ratio_1011_0011}
\begin{split}
 & \frac{M_\mathrm{v}^{(1011)}(t_1)}{M_\mathrm{v}^{(0011)}(t_1)}  \approx \frac{M_\mathrm{v0}^{(1011)}}{M_\mathrm{v}^{(0011)}(t_1)}  \left(\frac{M_\mathrm{v}^{(0011)}(t_1)}{M_\mathrm{v0}^{(0011)}}\right)^{\Gamma_{1011}/\Gamma_{0011}}\\
 & \approx 2.37 \times 10^{-3} \left(\frac{10^{-2} M_0}{M_\mathrm{v}^{(0011)}(t_1)}\right)^{7/8} 
\cdot \frac{{M_\mathrm{v0}^{(1011)}}}{{M_\mathrm{v0}^{(0011)}}} \left(\frac{M_\mathrm{v0}^{(0011)}}{10^{-5}M_0}\right)^{7/8},
   \end{split}
\end{align}
where $\Gamma_{1011}/\Gamma_{0011} \sim \beta_{10}/\beta_{00}=1/8$ with
\begin{align} 
    \beta_{nl} = \frac{(n+2l+1)!}{(n+l+1)^{2l+4}n!}.
\end{align}
It means the mass ratio of the subdominant mode to the dominant mode at $t_1$ is smaller if the latter is larger in size. This ratio is important for the modulation of the GW emission in \eref{eq:dE/dt_beat}, which will be discussed in Sec.~\ref{sec:detec}.

Since the subdominant mode has a much smaller superradiance rate and terminates earlier, this ratio is always small as long as the initial mass of the subdominant mode is not unnaturally large. Indeed, numerical calculation shows this ratio is as small as $10^{-3}$ with initial masses ${M_\mathrm{v0}^{(0011)}}={M_\mathrm{v0}^{(1011)}}=10^{-5}M_0$ and initial BH spin $a_{*0}=0.99$. In this work, we assume this ratio is always much smaller than identity.

During this time, the BH mass and spin drop rapidly to $M_1$ and $a_{*\text{c}}^{(1011)}$, respectively. Since the subdominant mode is much smaller in mass than the dominant one, one could make estimates of $M_1$ and $M_{\text{v}}^{0011}(t_1)$ as if only the dominant mode exists. Then the problem reduces to the single-mode evolution explained in Sec.~\ref{subsec:single_analytic}. $M_\text{v,max}^{(0011)}$ and $a_{*\text{c}}^{(0011)}$ in Eqs.~\eqref{eq:e_am_conservation} should be replaced by $M_\text{v}^{(0011)}(t_1)$ and $a_{*\text{c}}^{(1011)}$, respectively. Nonetheless, using $a_{*\text{c}}^{(1011)}$ or $a_{*\text{c}}^{(0011)}$ here leads to an error only at order $\mathcal{O}\left((M\mu)^8\right)$. As a result, Eq.~\eqref{eq:Mv_0011_max} can be used to estimate $M_\text{v}^{(0011)}(t_1)$.

With $M_\text{v}^{(0011)}(t_1)$ calculated, one could further obtain $M_\text{v}^{(1011)}(t_1)$ with Eq.~\eqref{eq:Max_ratio_1011_0011} and obtain $t_1$ with Eq.~\eqref{eq:two_modes_exp}. Apparently, $M_\text{v}^{(1011)}(t_1)$ is the maximum mass of the subdominant mode. The BH mass at time $t_1$ is then
\begin{align}
M_1 &\approx M_0 - {M_\mathrm{v}^{(1011)}}(t_1) - {M_\mathrm{v}^{(0011)}}(t_1). \label{eq:M_1}
\end{align}

The second phase starts at $t_1$ and ends at $t_2$ when the subdominant mode is drained. In this phase, the BH spin is between $a_{*\text{c}}^{(1011)}$ and $a_{*\text{c}}^{(0011)}$. The subdominant mode gradually falls into the BH while the dominant mode still extracts energy out of the horizon. At $t_2$, the BH spin equals to $a_{*\text{c}}^{(0011)}$. In this phase, the BH mass is almost unchanged, meaning the balance of the inflow and the outflow at the horizon
\begin{align}\label{eq:sub_attractor}
M_\mathrm{v}^{(0011)}(t)\Gamma_{0011}+M_\mathrm{v}^{(1011)}(t)\Gamma_{1011}=0.
\end{align}
Since $M_\mathrm{v}^{(0011)}(t) \gg M_\mathrm{v}^{(1011)}(t)$, the BH spin in this time range is very close to $a_{*\text{c}}^{(0011)}$. As an approximation, we assume the BH spin is always $a_{*\text{c}}^{(0011)}$ in this time range. Then the superradiance rate $\Gamma_{0011}=0$, indicating the horizon angular speed $\Omega_H=\omega_{0011}$. Use Eq.~\eqref{eq:SR_rate_Baumann} for the subdominant mode, one obtains
\begin{align}\label{eq:Gamma_1011_1}
\dot{E}_\text{H}^{(1011)}=2M_\text{v}^{(1011)} 
\Gamma_{1011} \approx -\frac{3}{4}\frac{M_\text{v}^{(1011)}}{M_1} (M_1\mu)^{10}.
\end{align}
On the other hand, the GW emission rate of the subdominant mode in Eq.~\eqref{eq:GW_single_1011_new} gives
\begin{align}
\dot{E}^{(1011)}_\mathrm{GW} 
\approx 6.79 \,a_{*0}
\frac{M_\text{v}^{(1011)}}{M_1} (M_1\mu)^{11},
\end{align}
where Eq.~\eqref{eq:Mv_0011_max} is used for $M_\text{v}^{(0011)}$. Compared to $\dot{E}_\text{H}^{(1011)}$, the GW emission rate $\dot{E}^{(1011)}_\mathrm{GW}$ has one more power of $M\mu$, but the overall coefficient is larger by a factor of 9. Inserting both terms into Eq.~\eqref{eq:Evoluton_E}, we see that after $t_1$, the mass $M_\text{v}^{(1011)}$ decreases exponentially as
\begin{align}
M_\text{v}^{(1011)}(t) = M_\text{v}^{(1011)}(t_1) e^{-(t-t_1)/\tau_\text{life}^{(1011)}},
\end{align}
where
\begin{align}\label{eq:life_1011}
\tau_\text{life}^{(1011)} = \frac{M_1}{(0.75+6.79\,a_{*0} M_1\mu) (M_1\mu)^{10}},
\end{align}
which could be used to estimate the lifetime of the subdominant mode. For the order of magnitude, one could take the $M\mu\to 0$ limit and obtain
\begin{align}\label{eq:1011_life}
    \tau_{\mathrm{life}}^{(1011)}\approx7.6\ \mathrm{days}\left(\frac{M_{0}}{10M_{\odot}}\right)\left(\frac{0.1}{M_{0}\mu}\right)^{10}.
\end{align}
 
Next, we study the dominant mode in the second phase. Combining Eqs.~\eqref{eq:dEH_dt}, \eqref{eq:sub_attractor} and \eqref{eq:Gamma_1011_1}, one could obtain
\begin{align}
\dot{E}_\text{H}^{(0011)} = 2 M_{\text{v}}^{(0011)} \Gamma_{0011} \approx \frac{3}{4}\frac{M_\text{v}^{(1011)}}{M_1} (M_1\mu)^{10}
\end{align}
The GW emission rate $\dot{E}^{(0011)}_\mathrm{GW}$ has been calculated in Eq.~\eqref{eq:GW_single}. The ratio of the two rates is
\begin{align}\label{eq:sr_GW_ratio}
\frac{\dot{E}_\text{H}^{(0011)}}{\dot{E}^{(0011)}_\mathrm{GW}}
\approx 0.03 \frac{M_1}{M_\mathrm{v}^{(0011)}(t)}  
\frac{M_\mathrm{v}^{(1011)}(t)}{M_\mathrm{v}^{(0011)}(t)}. 
\end{align}
This ratio at $t_1$ is always much less than 1. For other values of $t\in (t_1,t_2)$, the value is even smaller since the last fraction on the right side decreases with time exponentially. As a consequence, one could only consider the GW emission of the dominant mode evolution in this phase.

The vector and scalar superradiance have very different behavior in this phase. Instead of the exponential decay, the scalar dominant mode experiences a second superradiant growth \cite{Guo:2022mpr}.\footnote{Ref.~\cite{Guo:2022mpr} used the single-mode GW emission rate, the counterpart of Eq.~\eqref{eq:GW_single}, for the subdominant mode, while the strongest GW emission comes from the interference term, the counterpart of Eq.~\eqref{eq:GW_single_1011_new}. The qualitative behavior does not change after the correction.} 
This is attributed to the fact that in the scalar case, the GW emission rate is $\sim(M_1\mu)^4$ smaller, while the suppression of the superradiance rate is only $\sim(M_1\mu)^2$. As a result, the ratio in Eq.~\eqref{eq:sr_GW_ratio} is enhanced by $(M_1\mu)^{-2}$ for scalars. The ratio turns out to be larger than 1, causing a second superradiant growth of the scalar dominant mode.

The third phase starts from $t_2$ and ends at the time when the $j=2$ modes become important. The BH spin is slightly below $a_{*\text{c}}^{(0011)}$ throughout this phase. The dominant mode with $j=1$ shrinks. Most of its energy is emitted as GWs, while a small fraction is returned to the BH. The BH mass is at first $M_1$ but then presents a tiny local peak structure. The local peak forms because a vector in the $j=2$ modes takes twice the angular momentum as a vector in the $j=1$ modes. When the energy transfers from the $j=1$ dominant mode to the $j=2$ modes, some energy must be returned to the BH to keep the constant BH spin. We name the time of the local peak as $t_3$ and use it as the end of the third phase.

The dominant mode reaches its maximum mass at time $t_1$. Since then, its evolution is controlled by the GW emission since $t_1$, which is the same as the single-mode case explained in in Sec.~\ref{subsec:single_analytic}. Its mass can be estimated with Eq.~\eqref{eq:single-decay}. Its lifetime can be approximated by $\tau_\mathrm{GW}^{(0011)}$ in Eq.~\eqref{eq:tau_GW_0011}.

Beyond $t_3$, the BH-condensate system is in the $j=2$ stage.

\subsection{The $j=2$ stage}

The $j=2$ modes become important around $t_3$, when the dominant $j=1$ mode is almost drained. The evolution of the $j=2$ modes is very similar to that of the $j=1$ modes. There are also three phases. The first phase starts at $t=0$ and ends at $t_1'$, when the $^3P_{2,2}^1$ mode reaches its maximum mass. From the scaling at the beginning of this section, $t_1'$ is much larger than $t_3$. During this time, both $j=2$ modes grow exponentially. Since the BH mass and spin are $M_1$ and $a_{*\text{c}}^{(0011)}$ for the most time in this phase, these values should be taken as the ``initial" condition of the $j=m=2$ modes.

Analogous to the $j=1$ stage, we derive the maximum masses for the dominant $j=2$ mode at $t_1'$,
\begin{align}
    & \frac{M_\mathrm{v}^{(0122)}(t_1')}{M_1} \approx
    \frac{M_\mathrm{v0}^{(0122)}}{M_1}+ \left(M_1\mu\right)^2 - 6 \left(M_1\mu\right)^4,
    \label{eq:Mv_max_0122}
\end{align}
and the mass ratio of the subdominant $j=2$ mode to the dominant one,
\begin{align}
   \begin{split}
       & \frac{M_\mathrm{v}^{(1122)}(t_1')}{M_\mathrm{v}^{(0122)}(t_1')}  \approx \frac{M_\mathrm{v0}^{(1122)}}{M_\mathrm{v}^{(0122)}(t_1')}  \left(\frac{M_\mathrm{v}^{(0122)}(t_1')}{M_\mathrm{v0}^{(0122)}}\right)^{0.35}\\
       & = 1.12 \times 10^{-2}  \frac{{M_\mathrm{v0}^{(1122)}}}{{M_\mathrm{v0}^{(0122)}}} \left(\frac{10^{-2} M_0}{M_\mathrm{v}^{(0122)}(t_1')}\right)^{0.65} \left(\frac{M_\mathrm{v0}^{(0122)}}{10^{-5}M_0}\right)^{0.65}.
   \end{split}
   \label{eq:Max_ratio_1122_0122}
\end{align}
From $t_3$ to $t_1'$, the BH mass drops from $M_1$ to
\begin{align}
\begin{split}
    M_2 &\approx M_1 + M_\mathrm{v0}^{(0122)} +M_\mathrm{v0}^{(1122)}\\
    &\hspace{1cm} - {M_\mathrm{v}^{(0122)}(t_1')} - {M_\mathrm{v}^{(1122)}(t_1')}.
    \label{eq:M_2}
\end{split}
\end{align}
Meanwhile, the BH spin drops rapidly from $a_{*\text{c}}^{(0011)}$ to $a_{*\text{c}}^{(1122)}$.

The time $t_1'$ can be estimated with Eq.~\eqref{eq:two_modes_exp} with $i=0122$,
\begin{align}\label{eq:tau_sr_0122}
    t_1' \approx \frac{1}{2\Gamma_{0122}} \log\frac{M_\mathrm{v}^{(0122)}(t_1')}{M_\mathrm{v0}^{(0122)}},
\end{align}
where we follow the same argument above Eq.~\eqref{eq:Mv_max_vs_t} and choose $\Gamma_{0122}$ as its value at $t_3$,
\begin{align}
    M_1 \Gamma_{0122} = 9.26\times10^{-3} (M_1\mu)^{12}.
\end{align}
For the order of magnitude, we could set the logarithm to 10 and replace $M_1$ by $M_0$, which gives
\begin{align}
    t_1'\sim 843.4\ \mathrm{yr}\left(\frac{M_{0}}{10M_{\odot}}\right)\left(\frac{0.1}{M_{0}\mu}\right)^{12}.
\end{align}

The second phase starts at $t_1'$ and ends at $t_2'$ when the $^3P_{2,2}^1$ mode is drained. In this phase, the BH spin is between $a_{*\text{c}}^{(1122)}$ and $a_{*\text{c}}^{(0122)}$.  At $t_2'$, the BH spin equals to $a_{*\text{c}}^{(0122)}$. In this phase, the BH mass is almost unchanged with value $M_2$. Following the same steps as in the $j=1$ stage, one could show $M_\text{v}^{(1122)}$ decreases exponentially as
\begin{align}
M_\text{v}^{(1122)}(t) = M_\text{v}^{(1122)}(t_1') e^{-(t-t_1')/\tau_\text{life}^{(1122)}},
\end{align}
where
\begin{align}\label{eq:life_1122}
\tau_\text{life}^{(1122)} =  \frac{1000\,M_2}{\left(0.45+7.9\,(M_2\mu)^2\right) (M_2\mu)^{14}},
\end{align}
For the order of magnitude, one could take the $M\mu\to 0$ limit and replace $M_1$ by $M_0$, which gives
\begin{align}
    \tau^{(1122)}_\mathrm{life}  
    & \approx 3.46\times10^{5}\ \mathrm{yr}\left(\frac{M_{0}}{10M_{\odot}}\right)\left(\frac{0.1}{M_{0}\mu}\right)^{14}.
\end{align}
It is larger than the lifetime of the $^3S_{1,1}^1$ mode in Eq.~\eqref{eq:1011_life} by a factor of $10^8$ for $M_0\mu=0.1$.

In this phase, the evolution of the $^3P_{2,2}^0$ mode is dominated by the GW emission, the same as the $j=1$ stage. Consequently, its mass can be estimated as
\begin{align}\label{eq:0122_life}
M_\mathrm{v}^{(0122)}(t) & =
\frac{M_\mathrm{v}^{(0122)}(t_1')}{1+(t-t_1')/\tau_\mathrm{GW}^{(0122)}},
\end{align} 
with
\begin{align}
    & \tau_\mathrm{GW}^{(0122)} = \frac{M_2^2}{M_\mathrm{v}^{(0122)}(t_1')\, C_\mathrm{GW}^{(0122)}(M_2\mu)^{14}},
    \label{eq:tau_GW_0122}
    \\
    & \approx1.38\times10^{5}\ \mathrm{yr}\left(\frac{M_{0}}{10M_{\odot}}\right)\left(\frac{0.1}{M_{0}\mu}\right)^{14}\left(\frac{0.1}{{M_{\mathrm{v}}^{(0122)}}(t_1')/M_{0}}\right),
\end{align}
which is 7 orders of magnitude larger than $\tau_{\text{GW}}^{(0011)}$ for $M_0\mu=0.1$.

The third phase starts from $t_2'$ and ends at the time when the $j=3$ modes become important. The BH spin is slightly below $a_{*\text{c}}^{(0122)}$ throughout this phase. The $^3P_{2,2}^0$ mode shrinks. Its mass is still described by Eq.~\eqref{eq:0122_life}. Most of its energy is emitted as GWs, while a small fraction is returned to the BH. The BH mass is at first $M_2$ but then presents a tiny local peak structure. The local peak forms because a vector in the $j=3$ modes takes 1.5 times the angular momentum as a vector in the $j=2$ modes. When the energy transfers from the $j=2$ dominant mode to the $j=3$ modes, some energy must be returned to the BH to keep the constant BH spin. We name the time of the local peak as $t_3'$ and use it as the end of the third phase.

Beyond $t_3'$, the BH-condensate system is in the $j=3$ stage. The evolution is very similar to the $j=1$ and $j=2$ stages.

Note that the above estimates assume the initial BH spin $a_{*0}$ to be larger than $a_{*\mathrm{c}}^{(0011)}$ such that the  $j=m=1$ modes are enhanced via superradiance. If the initial BH spin satisfies $a_{*\mathrm{c}}^{(1122)}<a_{*0}<a_{*\mathrm{c}}^{(0011)}$, the $j=2$ modes are the fastest-growing modes. Then one should estimate the maximum mass of the $^3P_{2,2}^0$ mode with,
\begin{align}\label{eq:Mv_max_0122_new}
    \begin{split}
        \frac{M_\mathrm{v}^{(0122)}(t_1')}{M_0} \approx & M_0\mu \Big[ \frac{a_{*0}}{2}- M_0\mu + 0.97 a_{*0} (M_0\mu)^2 \\
        & \hspace{1cm}+(-0.75 + 0.10 a_{*0}^2)(M_0\mu)^3\Big],
    \end{split}
\end{align}
instead of  \eref{eq:Mv_max_0122}. This equation is used in the calculation of Fig.~\ref{fig:Parameter_j=m=2} below.

\subsection{Entire evolution}

\begin{figure}[!ht]
    \centering
    \includegraphics[width=0.48\textwidth]{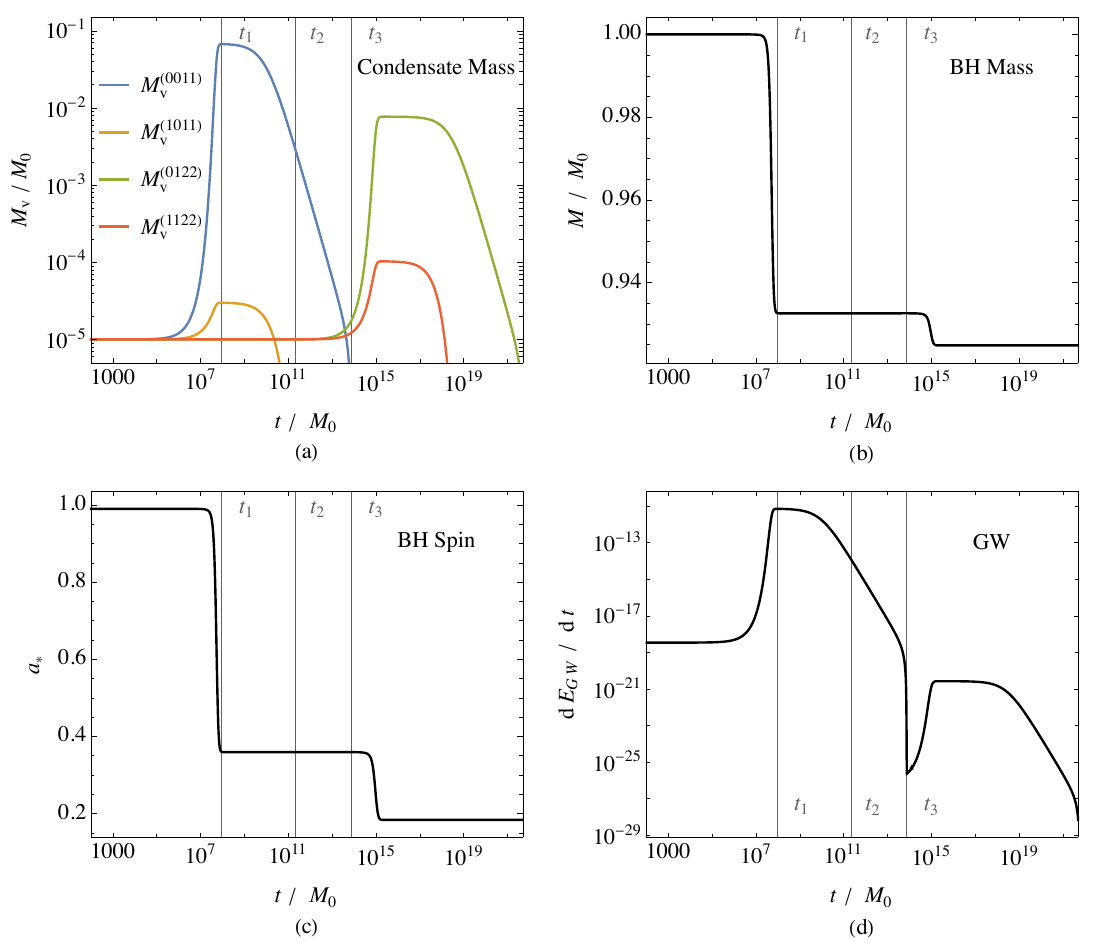}
    \caption{The evolutions of the vector condensate mass (panel a), BH mass (panel b), BH spin (panel c), and total GW emission flux (panel d) as functions of time. The initial parameters are $a_{*0}=0.99$, $M_0\mu=0.1$ and $M_\mathrm{v0}^{(nljm)}=10^{-5}M_0$ for each mode. Two dominant modes ($^3S_{1,1}^0$ and $^3P_{2,2}^0$) and two subdominant modes ($^3S_{1,1}^1$ and $^3P_{2,2}^1$) are included in the evolution. The three vertical lines label the three critical times (see text for details).}
    \label{fig:evolution_four_modes}
\end{figure}

In this part, we numerically solve the evolution equations in Eqs.~\eqref{eq:Evoluton_ODE} and compare the results with the analytic approximations obtained above. The initial BH masses and spin are set as $M_0\mu=0.1$ and $a_{*0}=0.99$, respectively. Two $j=1$ modes ($^3S_{1,1}^0$ and $^3S_{1,1}^1$) and two $j=2$ modes ($^3P_{2,2}^0$ and $^3P_{2,2}^1$) are included in the numerical calculation. The initial mass of the four modes are all set as $10^{-5}M_0$. The time dependence of the mass of each mode, the BH mass and spin, as well as the GW flux, are shown in Fig.~\ref{fig:evolution_four_modes}. The comparison of the analytic approximations and the numerical values are further exhibited in Table.~\ref{tab:four_modes}.

\begin{table*}
    \renewcommand\arraystretch{1.4}
    \begin{centering}
    \begin{tabular}{c|ccccccc}
    \hline 
    Quantities & $M_{\mathrm{v}}^{(0011)}(t_1)/M_{0}$ & $M_{\mathrm{v}}^{(1011)}(t_1)/M_{0}$ & $t_1/M_{0}$ & $\tau_{\mathrm{life}}^{(1011)}/M_{0}$ & $\tau_{\mathrm{GW}}^{(0011)}/M_{0}$ & $M_{1}/M_{0}$ & $a_{*1}$\tabularnewline
    \hline 
    Estimates & $6.59\times10^{-2}$ & $3.00\times10^{-5}$ & $4.79\times10^{7}$ & $1.34\times10^{10}$ & $9.64\times10^{9}$ & 0.934 & 0.359\tabularnewline
    Numerical results & $6.71\times10^{-2}$ & $2.99\times10^{-5}$ & $9.51\times10^{7}$ & $2.07\times10^{10}$ & $9.69\times10^{9}$ & 0.933 & 0.359\tabularnewline
    \hline 
    \hline 
    Quantities & $M_\mathrm{v}^{(0122)}(t_1')/M_{0}$ & $M_\mathrm{v}^{(1122)}(t_1')/M_{0}$ &  $t_1'/M_{0}$ & $\tau_{\mathrm{life}}^{(1122)}/M_{0}$ & $\tau_{\mathrm{GW}}^{(0122)}/M_{0}$ & $M_{2}/M_{0}$ & $a_{*2}$\tabularnewline
    \hline 
    Estimates & $7.73\times10^{-3}$ & $1.03\times10^{-4}$ & $8.13\times10^{14}$ & $5.22\times10^{17}$ & $2.87\times10^{18}$ & 0.926 & 0.183\tabularnewline
    Numerical results & $7.69\times10^{-3}$ & $1.03\times10^{-4}$ & $2.32\times10^{15}$ & $5.79\times10^{17}$ & $3.00\times10^{18}$ & 0.925 & 0.183\tabularnewline
    \hline 
    \end{tabular}
    \par\end{centering}
    \caption{Numerical results and estimates for the important quantities in the evolution of the BH-condensate system. The initial parameters are identical to those used in \figref{fig:evolution_four_modes}.}
    \label{tab:four_modes}
\end{table*}

In Fig.~\ref {fig:evolution_four_modes}, the numerical values of $t_1$, $t_2$, and $t_3$ are presented by vertical lines explicitly. It is clear that both $j=1$ and $j=2$ stages consist of three phases. Using the $j=1$ stage as an illustration, we name the three phases and describe the processes below.  The time is measured in the unit of $M_0$, which can be easily converted to the SI units with $M_\odot=4.92\times 10^{-6}$~s.

\begin{enumerate}
    \item \textbf{Spin-down phase:} In the time range $0<t\leqslant t_1$ with $t_1=9.51\times10^{7}M_0$, the BH spins drops to $0.359$ rapidly. The BH mass also drops slightly to $0.933M_0$. Meanwhile, both  the dominant $^3S_{1,1}^0$ mode and the subdominant $^3S_{1,1}^1$ mode grow exponentially, from $10^{-5}M_0$ to $6.71\times10^{-2}M_0$ and $2.99\times10^{-5}M_0$, respectively. These values are also their maximum masses in the entire evolution. The integrated GW emission energy in this period is $2.80\times10^{-4}M_0$. The growth of the two $j=m=2$ modes during this period is too small to be observable in \figref{fig:evolution_four_modes}.
    \item \textbf{Subdominant attractor phase:} In the time range $t_1<t\leqslant t_2$, with $t_2=2.23\times10^{10}M_0$, the subdominant $^3S_{1,1}^1$ mode shrinks exponentially until is drained at $t_2$. Numerically, we call a mode is drained when its mass is below $10^{-8}M_0$. The BH mass and spin remain roughly constant at $0.933M_0$ and $0.359$, respectively.  Mathematically, the BH is at the attractor because of the presence of the $^3S_{1,1}^1$ mode. Meanwhile, the dominant $^3S_{1,1}^0$ mode loses $6.435\times10^{-2}M_0$ in mass, and dissipates $6.442\times10^{-2}M_0$ via GW emission. The small discrepancy is due to the extraction of energy from the BH. It confirms the argument that the dominant mode in this phase is governed by the GW emission. In contrast, the subdominant $^3S_{1,1}^1$ mode loses $2.99\times10^{-5}M_0$ in mass, more than 3 times larger than the energy $7.23\times10^{-6}M_0$ emitted as GW. The energy difference is absorbed by the BH. Thus both the absorption and the GW emission are important for the evolution of the subdominant mode in this phase.
    \item \textbf{Dominant attractor phase:} In the time period $t_2<t\leqslant t_3$ with $t_3=7.43\times10^{13}M_0$, the BH spin is a constant slightly smaller than 0.359, while the BH mass increases very slowly from $0.932542M_0$ at $t_2$ to a local maximum value $0.932552M_0$ at $t_3$. The local peak is too small to be observed in Fig.~\ref{fig:evolution_four_modes}. In this time range, the dominant $^3S_{1,1}^0$ mode loses $2.77\times10^{-2}M_0$ in mass, while the dissipation via GW emission is  $2.75\times10^{-2}M_0$. The small energy difference is absorbed by the BH. This agrees with our analysis above which leads to the analytic approximations. 
\end{enumerate} 

After $t_3$, the $j=2$ stage starts and repeats these three phases. The GW emission flux in Fig.~\ref{fig:evolution_four_modes}(d) presents a sharp decline with $t\gtrsim t_3$. At this time, the energy transfers from the dominant $j=1$ mode to $j=2$ modes. Nonetheless, the GW emission of the latter is much smaller than the former, with $\dot{E}^{(0122)}_\mathrm{GW} \sim 10^{-4}(M\mu)^{4}\dot{E}^{(0011)}_\mathrm{GW}$, which explains the sharp decline of the GW flux close to $t_3$.

\begin{figure}[!ht]
    \centering
    \includegraphics[width=0.48\textwidth]{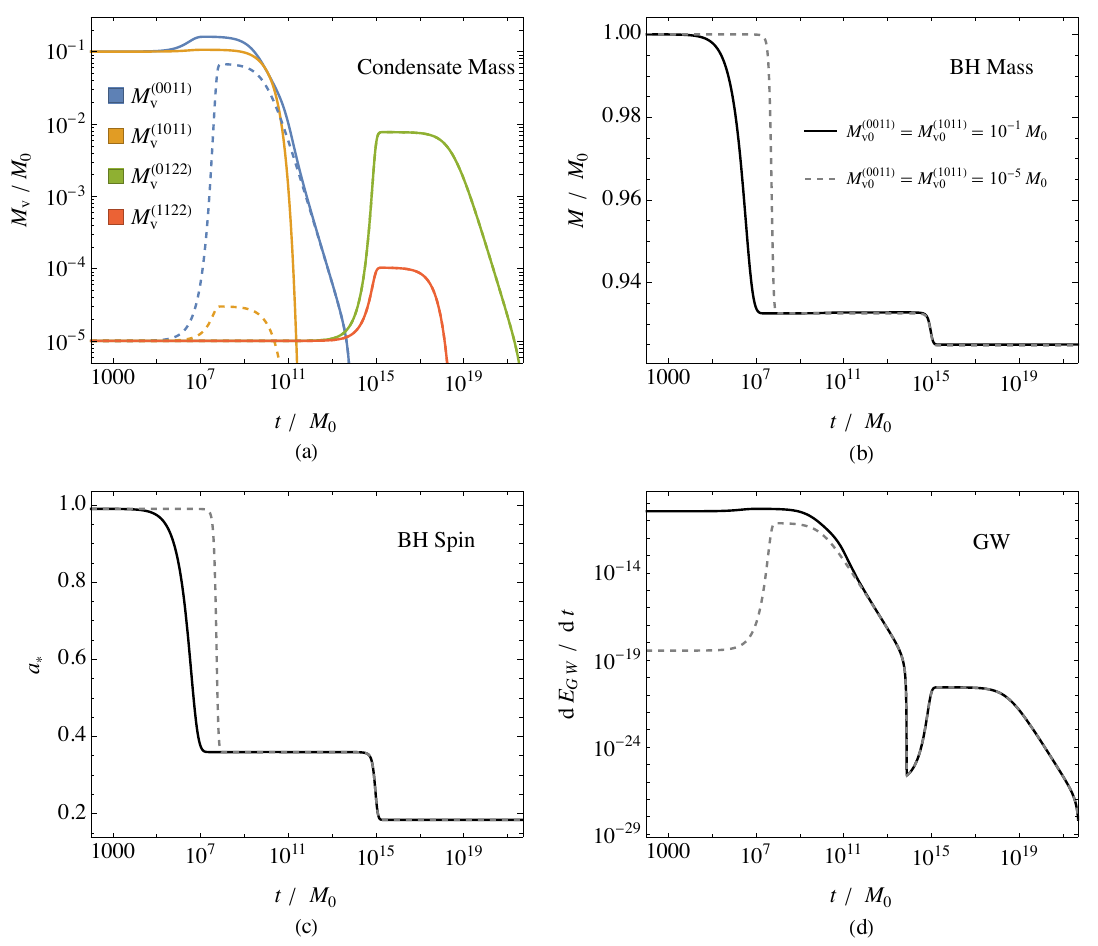}
    \caption{
The evolutions of the vector condensate mass (panel a), BH mass (panel b), BH spin (panel c), and total GW emission flux (panel d) as functions of time. Two dominant modes ($^3S_{1,1}^0$ and $^3P_{2,2}^0$) and two subdominant modes ($^3S_{1,1}^1$ and $^3P_{2,2}^1$) are included in the evolution. The initial parameters are $a_{*0}=0.99$ and $M_0\mu=0.1$. The initial masses of the $j=1$ modes are chosen as $10^{-5}M_0$ and  $10^{-1}M_0$. Numerical calculation confirms that the initial masses of small-$j$ modes have no impact on the large-$j$ modes, allowing the direct connection between the initial BH parameters and the large-$j$ mode evolution.}
    \label{fig:evolution_four_modes_2}
\end{figure}

Finally, we explore an interesting aspect of vector superradiance. In Sec.~\ref{subsec:Mv0}, we have found that neither the BH nor the $j=1$ mode at the end of the $j=1$ stage depends on the initial condensate mass. Since the BH mass and spin at that time serve as the initial condition of the $j=2$ stage, a natural conjecture is that the initial masses of the $j=1$ modes do not affect the evolution of the $j=2$ stage. In Fig.~\ref{fig:evolution_four_modes_2}, we recalculate the evolution of the four modes numerically with different parameters. The initial masses of the two $j=1$ modes are changed from $10^{-5}M_0$ to $10^{-1}M_0$, while all other parameters are the same as in Fig.~\ref{fig:evolution_four_modes}. As expected, the $j=2$ stages of these two cases are exactly the same. Indeed, the evolution of the $j=2$ modes only depends on the BH initial mass and spin, as well as the initial masses of the $j=2$ modes at $t=0$. This is important when we study the beat signature from the $j=2$ modes.

\section{The detection of vector superradiance}\label{sec:detec}

\begin{figure*}[htbp]
    \centering
    \includegraphics[width=0.8\textwidth]{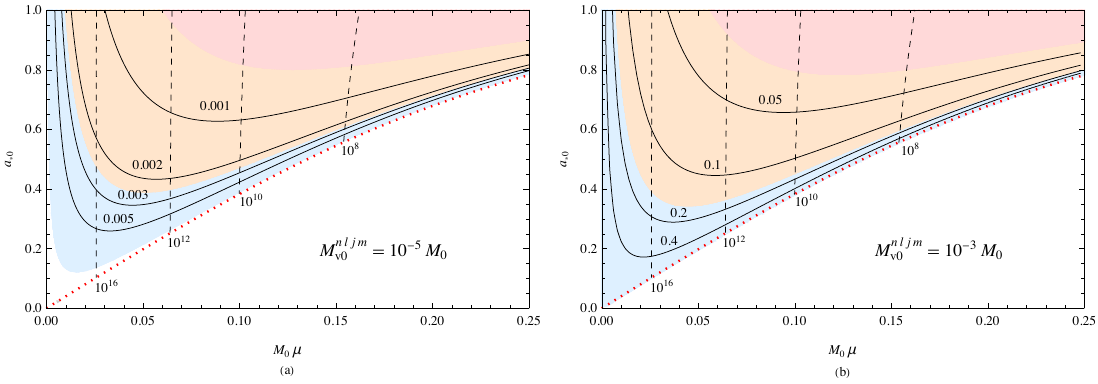}
    \caption{The strength and the duration of the GW beat with different BH initial parameters. The initial masses of the vector condensates are $M_{\text{v}0}^\text{(0011)} = M_{\text{v}0}^\text{(1011)}= 10^{-5}M_0$ (panel a) and $M_{\text{v}0}^\text{(0011)} = M_{\text{v}0}^\text{(1011)}= 10^{-3}M_0$ (panel b). The red dotted line is $a_{*0}=a^{(0011)}_\mathrm{*c}$, below which the $^3S_{1,1}^0$ superradiance does not exist. The blue, orange, and pink shaded regions are for $\left(M^{(0011)}_\mathrm{v,max}+M^{(1011)}_\mathrm{v,max}\right)/M_1$ being greater than 0.001, 0.01, 0.05, respectively. The solid contours are calculated with $M^{(1011)}_\mathrm{v,max}/M^{(0011)}_\mathrm{v,max}$ being the constants labelled above the curves. The dashed contours are for constant $\tau_\mathrm{life}^{(1011)}$ labeled beneath the curves, with values in the unit of $M_0$. The conversion from $M_0$ to the SI units is $M_\odot =4.92\times 10^{-6}$~s.}
    \label{fig:Parameter_j=m=1}
\end{figure*}
\begin{figure*}[ht]
    \centering
    \includegraphics[width=0.8\textwidth]{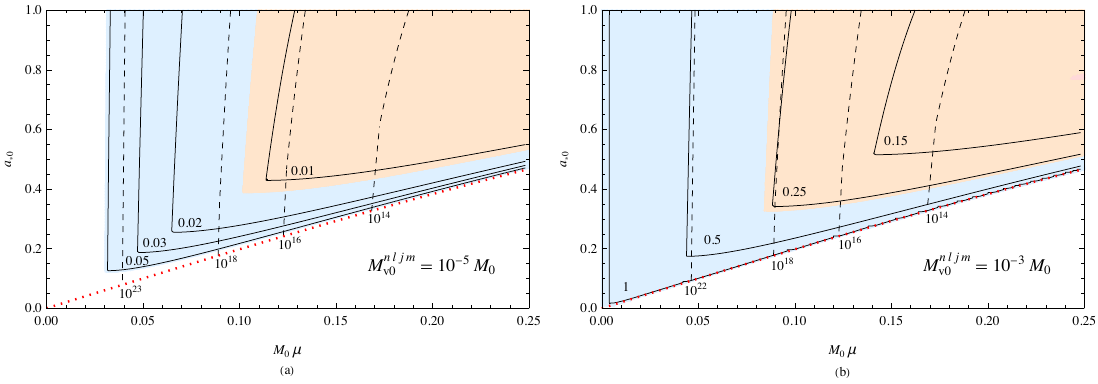}
    \caption{Same as \figref{fig:Parameter_j=m=1}, but for $j=m=2$ modes. }
    \label{fig:Parameter_j=m=2}
\end{figure*}

Detecting vector superradiance can be either directly through the GW signals, or indirectly by analyzing the observed BH mass and spin distribution.
In this section, we study both methods in detail.

\subsection{Gravitational wave signals}

Although a Kerr BH does not radiate GW, the rotating vector condensate does. With only the dominant mode, the emitted GW is monochromatic, with a frequency equal to twice the energy of a vector particle in that mode. Nonetheless, there are other candidates, such as neutron stars, that could radiate monochromatic GW. Including the subdominant modes could break the degeneracy. Because the dominant and subdominant modes have a small difference in frequency, the coexistence of both modes results in a unique beat signature in the GW. Ref.~\cite{Guo:2022mpr} studied the beat signal from the BH-scalar-condensate system. In this section, we study similar beat signals from the BH-vector-condensate systems. 

There are three quantities essential for judging whether the beat signal is observable: (1) the mass of the condensate determining the strength of the GW, described by Eqs.~\eqref{eq:Mv_0011_max}, \eqref{eq:Max_ratio_1011_0011}, \eqref{eq:Mv_max_0122} and \eqref{eq:Max_ratio_1122_0122}; (2) the mass ratio of the subdominant mode to the dominant mode determining the strength of the GW beat, described by Eqs.~\eqref{eq:Max_ratio_1011_0011} and \eqref{eq:Max_ratio_1122_0122}; (3) the lifetime of the subdominant mode determining the duration of the GW beat, described by Eqs.~\eqref{eq:life_1011} and \eqref{eq:life_1122}. Notably, this duration is comparable to the timescale of the GW emitted by the dominant mode. These quantities for the $j=1$ stage with different initial parameters are shown in \figref{fig:Parameter_j=m=1}. The durations are labeled in the unit of $M_0$, which can be converted to seconds using $M_\odot = 4.92\times 10^{-6}$~s. The vector mass $\mu$ in unit eV can be obtained with the dimensionless $M_0\mu$ and $M_\odot = 7.5\times 10^9$~eV$^{-1}$.

We assume the BH initial spin is greater than $a_{*\text{c}}^{(1011)}$, such that the $j=1$ modes grow via superradiance at first. The GW beat signal is determined by the BH initial mass and spin, as well as the initial masses of the $j=1$ modes. In Fig.~\ref{fig:Parameter_j=m=1}, we show the properties of the GW beat signal as functions of the initial parameters. The GW radiated by the $j=2$ modes becomes important only after the $j=1$ modes are drained. One may suspect the evolution of the $j=1$ modes complicates the relation between the BH initial parameters and the beat signal in the $j=2$ stage. This conjecture is nevertheless incorrect from our calculation in Sec.~\ref{sec:Multi-mode}. In particular, the initial masses of the $j=1$ modes do not affect the evolution of the $j=2$ modes at all, as illustrated in Fig.~\ref{fig:evolution_four_modes_2}. As a result, it is qualified to relate the properties of the GW beat signal from the $j=2$ modes to the initial BH mass and spin in Fig.~\ref{fig:Parameter_j=m=2}.

Since the GW signal is unavoidably mixed with the stochastic background, either large strength or more features are necessary to separate the signal. In both Figs.~\ref{fig:Parameter_j=m=1} and \ref{fig:Parameter_j=m=2}, a strong GW requires both the BH initial BH mass and spin to be large. The rest of the parameter space can be explored with the help of the beat signal. Specifically, the strength of the beat is anti-correlated to the strength of the GW, which is expected with Eqs.~\eqref{eq:Max_ratio_1011_0011} and \eqref{eq:Max_ratio_1122_0122}. Therefore, the synergy of the GW strength and the beat feature could be important in searching for the BH-condensate systems in experiments.

In Fig.~\ref{fig:Parameter_j=m=1}, one finds the $j=m=1$ modes radiate strong GWs in a rather short time. It is proposed to observe the GW signal in the $j=1$ stage modes right after two BHs merge~\cite{Siemonsen:2022yyf}. Comparably, the $j=2$ stage has a much longer GW emission time with the same parameters, longer than the $j=1$ stage by more than 7 orders of magnitude. The GW in the $j=2$ stage also has a much stronger beat signal, resulting from the larger mass ratio of the subdominant mode to the dominant one. Although the GW strength is weaker, the longer duration and stronger beat signal of the GW in the $j=2$ stage could facilitate the observation, providing a complementary method to search for ultralight vectors in the BH merger events.

For discussing detectability, the GW emission flux \eqref{eq:dE/dt_beat} should be converted to the GW strain amplitude. The details are available in Refs.~\cite{Brito:2017zvb,Brito:2017wnc,Brito:2020lup,Guo:2022mpr}. We first ignore the frequency differences in \eref{eq:dE/dt_beat}. These quasi-monochromatic GWs have a frequency of $f_\mathrm{s} = \widetilde{\omega}/(2\pi) \simeq \mu/\pi$ in the source frame. The corresponding period is given by
\begin{align}
    T_\mathrm{s,GW}\approx21\ \mathrm{sec} \left(\frac{10^{-16}\mathrm{eV}}{\mu}\right).
\end{align}
In this case, the flux and amplitude have a simple relation \cite{Brito:2017zvb},
\begin{align}
    h_\mathrm{c}=\sqrt{\frac{N_{\text{cycles}}}{5f_\mathrm{s}^2r^2\pi^2}\frac{dE_\text{GW}}{dt}},
    \label{eq:hc_vs_flux}
\end{align}
where $r$ is the comoving distance. In the detector frame, the GW frequency is redshifted to $f_\mathrm{d} \equiv f_\mathrm{s}/(1+z)$. One could simply replace $f_\text{s}r$ in the formula by $f_\text{d} r_\text{L}$, where $r_\text{L}=r(1+z)$ is the luminosity distance. $N_{\text{cycles}}$ represents the number of observed cycles, which is approximated by the minimum of the $f_\mathrm{d}T_\mathrm{obs}$ and $f_\mathrm{s}\tau_\mathrm{GW}$ in our calculation, where $T_\mathrm{obs}$ is the observation time. If the lifetime is short, $N_{\text{cycles}}$ is determined by the source lifetime; otherwise, it is constrained by the design of GW interferometers.

Inserting \eref{eq:GW_single} into \eref{eq:hc_vs_flux}, one obtains a general expression for dominant modes,
\begin{align}
    h_\mathrm{c}^{(0(j-1)jj)}= \sqrt{\frac{N_{\text{cycles}}C_{\mathrm{GW}}^{(0(j-1)jj)}}{5}}\frac{M}{r}\frac{M_\mathrm{v,max}^{(0(j-1)jj)}}{M}(M\mu)^{2l+4}.
    \label{eq:hc_single}
\end{align}
For the $^3S_{1,1}^0$ and $^3P_{2,2}^0$ modes, one arrive at
\begin{align}
    \begin{split}
        h_{\mathrm{c}}^{(0011)} & = 1.12\times10^{-24}\\
        &\hspace{0.5cm}\times\sqrt{N_{\text{cycles}}}\frac{\mathrm{Mpc}}{r}\frac{M}{10M_{\odot}}\frac{a_{*0}-4M\mu}{0.1}(M\mu)^{5},
    \end{split} \label{eq:hc_0011}
    \\
    h_{\mathrm{c}}^{(0122)} & = 2.27\times10^{-28}\sqrt{N_{\text{cycles}}}\frac{\mathrm{Mpc}}{r}\frac{M}{10M_{\odot}}(M\mu)^{8}.
    \label{eq:hc_0122}
\end{align}
Choosing an observation time $T_\mathrm{obs} = 4\, \text{yr}$, a vector mass $10^{-16} \text{eV}$, and a redshift $z = 1$, the number of observed cycles $N_{\text{cycles}}$ is approximately $10^6$.

When a dominant mode and a subdominant mode coexist, there is a characteristic beat signal due to the small frequency difference $\Delta\widetilde{\omega}$. The period of the beat in the source frame can be expressed as $T_\mathrm{s,beat}={2\pi}/{\Delta\widetilde{\omega}}$. Specifically, for the $j=m=1$ modes and $j=m=2$ modes, the periods are
\begin{align}
    \begin{split}
        T^{m=1}_\mathrm{s,beat} & \approx 533.3  \left(\frac{0.1}{M\mu}\right)^2 T_\text{GW}\\
        &\approx 3.06 \, \mathrm{h} \left(\frac{10^{-16}\text{eV}}{\mu}\right) \left(\frac{0.1}{M\mu}\right)^2,
    \end{split}
    \\
    \begin{split}
        T^{m=2}_\mathrm{s,beat} & \approx 2880  \left(\frac{0.1}{M\mu}\right)^2 T_\text{GW}\\
        &\approx 16.5 \, \mathrm{h} \left(\frac{10^{-16}\text{eV}}{\mu}\right) \left(\frac{0.1}{M\mu}\right)^2.
    \end{split}
\end{align}
In this case, the characteristic strain amplitude becomes \cite{Guo:2022mpr}
\begin{align}\label{eq:hc_mod}
    h_\mathrm{c}= & \sqrt{\frac{4N_{\text {cycles }}}{5\mu^2r^2}\cdot 
    \left(\mathfrak{h}^\mathrm{ave}+\mathfrak{h}^\mathrm{mod}\right)
    },
\end{align}
where the sum $\mathfrak{h}^\mathrm{ave}+\mathfrak{h}^\mathrm{mod}$ can be derived by changing every occurrence of $\left|U^{(\widetilde{\omega}_i)}_{\widetilde{l}\widetilde{m}}\right|$ in \eref{eq:dE/dt_beat} to  $\left|U^{(\widetilde{\omega}_i)}_{\widetilde{l}\widetilde{m}}\right|/\widetilde{\omega}_i^2$ with $i=1,2,3$. $\mathfrak{h}^\mathrm{ave}$ is independent of time, while $\mathfrak{h}^\mathrm{mod}$ is time-modulated. If all sources have the same frequency, the modulated term $\mathfrak{h}^\mathrm{mod}$ is absent and Eq.~\eqref{eq:hc_mod} reduces to \eref{eq:hc_vs_flux}. Moreover, if the particle number of the dominant mode $N_d$ is much larger than that of the subdominant mode $N_s$, $\mathfrak{h}^\mathrm{ave}$ and $\mathfrak{h}^\mathrm{mod}$ could be further simplified,
\begin{align}
    \mathfrak{h}^\mathrm{ave} &\approx \frac{1}{8\pi}\dfrac{N_{d}^{2}}{{\omega_{d}}^2} \frac{\left|U_{\widetilde{l}\widetilde{m}}^{(\widetilde{\omega}_1)}\right|^{2}}{\widetilde{\omega}_{1}^4},\\
    \begin{split}
        \mathfrak{h}^\mathrm{mod} &\approx \frac{1}{2\pi} \sqrt{\dfrac{N_{d}^3N_{s}}{{\omega_{d}}^3\omega_{s}}} \frac{\left|U_{\widetilde{l}\widetilde{m}}^{(\widetilde{\omega}_1)}\right| \left|U_{\widetilde{l}\widetilde{m}}^{(\widetilde{\omega}_3)}\right|}{\widetilde{\omega}_{1}^2 \widetilde{\omega}_{3}^2}\\
        &\hspace{1cm}\times\cos\left[\widetilde{\omega}_4\left(t-r\right)
        -\phi_{\widetilde{l}\widetilde{m}}^{(\widetilde{\omega}_3)}+\phi_{\widetilde{l}\widetilde{m}}^{(\widetilde{\omega}_1)}\right].
    \end{split}
\end{align}
We define the average characteristic strain as $\bar{h}_\mathrm{c} \equiv \sqrt{4N_{\text {cycles }}\mathfrak{h}^\mathrm{ave}/(5\mu^2r^2)}$ and the beat characteristic strain as $\delta{h}_\mathrm{c}\equiv \max{(h_\mathrm{c})}-\min{(h_\mathrm{c})}$. In the limit of $N_d\gg N_s$, $\bar{h}_c$ reduces to \eref{eq:hc_single} and
\begin{align}
    \delta{h}_\mathrm{c}^{(nljm)} \approx 4\sqrt{\dfrac{N_{s}}{N_{d}}}\frac{\left|U_{\widetilde{l}\widetilde{m}}^{(\widetilde{\omega}_{3})}\right|}{\left|U_{\widetilde{l}\widetilde{m}}^{(\widetilde{\omega}_{1})}\right|}\cdot \bar{h}_\mathrm{c}^{(nljm)}.
    \label{eq:hc_beat}
\end{align}
The fraction ${\left|U_{\widetilde{l}\widetilde{m}}^{(\widetilde{\omega}_{3})}\right|}/{\left|U_{\widetilde{l}\widetilde{m}}^{(\widetilde{\omega}_{1})}\right|}$ is $\sqrt{2}/4\approx0.35$ for the $j=1$ stage and $16/27\approx0.59$ for the $j=2$ stage. Thus, the beat modulation could reach $5\%$ even when $N_{s}/N_{d}$ is as small as $10^{-3}$.

\begin{figure}[htbp]
    \centering
    \includegraphics[width=0.45\textwidth]{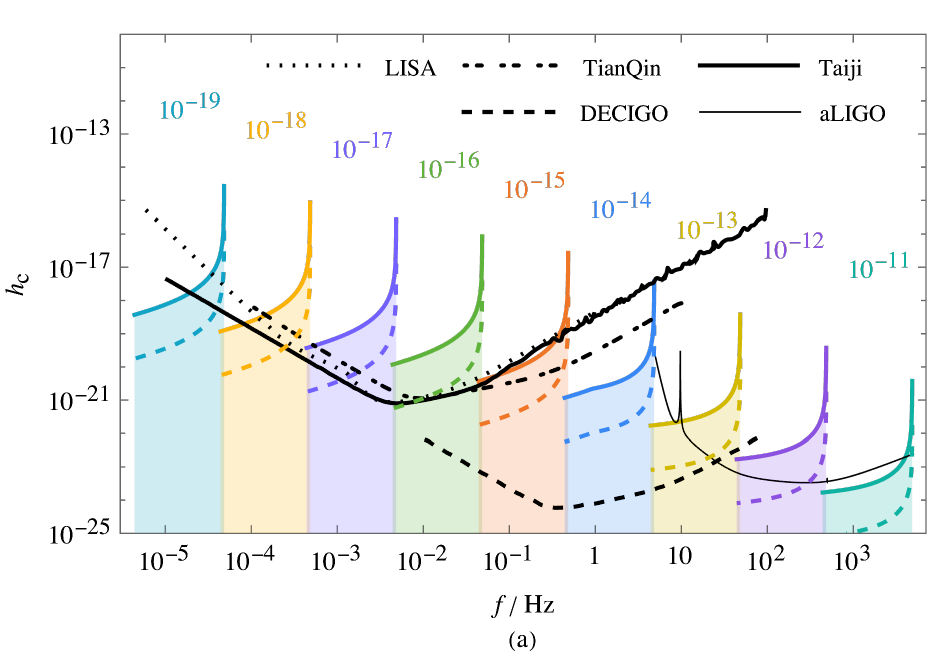}
    \includegraphics[width=0.45\textwidth]{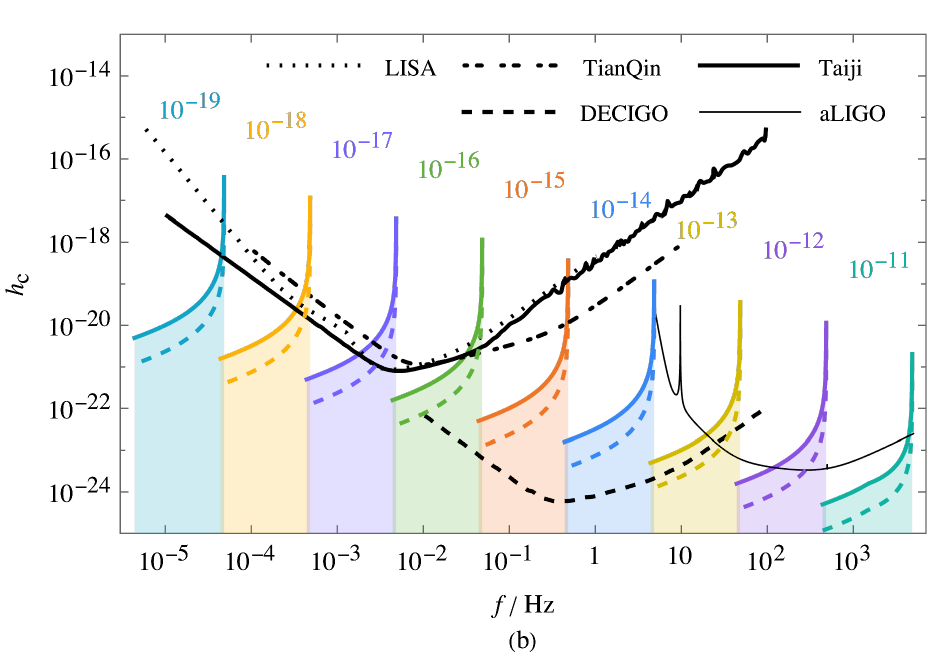}
    \caption{    
   \textbf{Panel (a):} The average characteristic strain $\bar{h}_c$ (solid colored curves) and the size of modulation $\delta h_c$ (dashed colored curves) from $j=1$ modes, together with the characteristic noise strain (black curves) of LISA \cite{LISA:2017pwj,Robson:2018ifk}, TianQin  \cite{TianQin:2015yph}, Taiji \cite{Luo:2021Manual}, DECIGO \cite{Kawamura:2006up}, and Advanced LIGO  \cite{Barsotti:T1800044v5}. The horizontal coordinate $f$ is the GW frequency in the detector frame. Each color band represents a vector mass labelled above the band in the unit of eV. From left to right on each band, the redshift $z$ decreases from 10 to 0.001. The parameters are $\alpha_{*}=0.1$, $M^\mathrm{(0011)}_\mathrm{v}+M^\mathrm{(1011)}_\mathrm{v}=0.1M$ and $N_{s}/N_{d}=10^{-3}$. The observation time is the minimum of 4~years and the signal lifetime in the detector frame.
\textbf{Panel (b):} Similar to panel (a) but for the $j=2$ modes. The parameters are $\alpha_{*}=0.25$, $M^\mathrm{(0122)}_\mathrm{v}+M^\mathrm{(1122)}_\mathrm{v}=0.01M$ and $N_{s}/N_{d}=10^{-2}$.}
    \label{fig:GW_hc}
\end{figure}

The average and the modulated characteristic strains are plotted in \figref{fig:GW_hc}, together with the characteristic noise strains of the current and projected GW detectors. For the $j=1$ stage, we assign the total condensate mass as 10\% of the BH mass, consistent with \figref{fig:Parameter_j=m=1}. The occupancy number ratio of the subdominant mode to the dominant mode is set as $N_{s}/N_{d}=10^{-3}$. Different colors correspond to various vector masses, as indicated by the numbers in the unit of eV associated with each color. The solid line atop the shaded areas represents the initial mass coupling given by $\alpha_0=0.1$. For the $m=2$ stage, we choose the total condensate mass as 1\% of the BH initial mass. The mass ratio of the subdominant mode to the dominant mode is $N_{s}/N_{d}=10^{-2}$ and the initial mass coupling is $\alpha_\mathrm{0}=0.25$.

From \figref{fig:GW_hc}, both the monochromatic GW and the GW beat in the $j=1$ stage have good detectability for DECi-hertz Interferometer Gravitational wave Observatory (DECIGO) (from $10^{-16}$ to $10^{-13}$ eV), advanced LIGO (from $10^{-13}$ to $10^{-12}$ eV) and space-based detectors (from $10^{-19}$ to $10^{-15}$ eV). In the $j=2$ stage, DECIGO still has a very good potential while other detectors can only search for sources in our neighborhood. It is worth noting that the parameters in the above calculation are rather conservative. The detectability would be greatly improved with a larger initial BH mass and/or condensate mass.

\subsection{Black hole Regge trajectories}\label{sec:indirect}

In addition to the direct detection of vector superradiance via gravitational waves, the distribution of BH spin and mass offers an indirect yet powerful method for detection. As previously discussed, a high-spin BH 
reduces its spin rapidly in the spin-down phase of each stage of evolution, eventually aligning with the Regge trajectory of the dominant modes, i.e., the $^3(j-1)_{j,j}^0$ modes. This alignment results in large forbidden regions in the BH "Regge plot"—a plane that maps spin against mass. By scrutinizing the data from numerous BHs, we could identify both favored and unfavored vector mass ranges.

\begin{figure*}[ht]
    \centering
    \includegraphics[width=\textwidth]{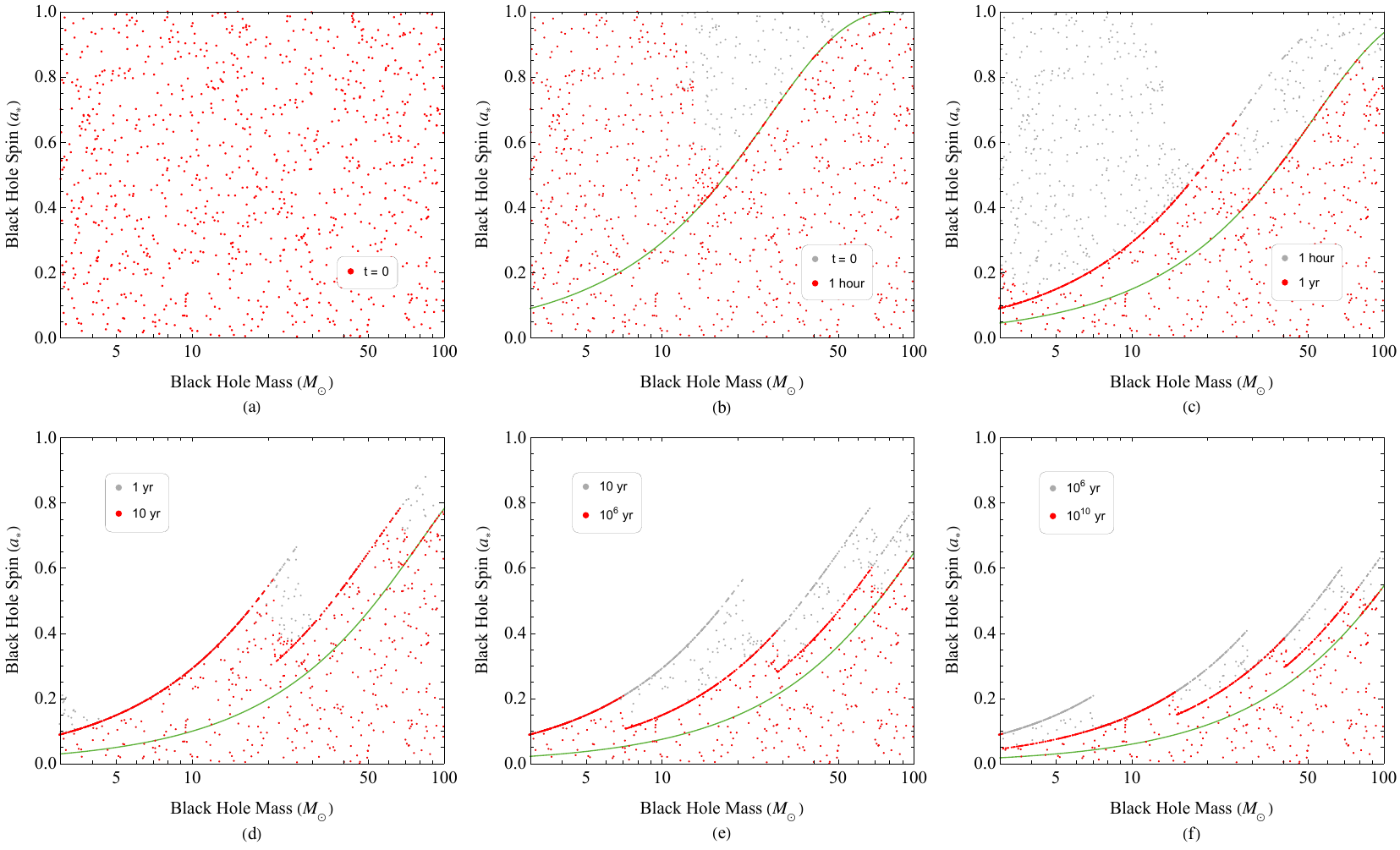}
    \caption{Evolution of BH spin-mass distribution with vector superradiance from solving Eqs.~\eqref{eq:Evoluton_ODE}.
    The vector mass is fixed at $10^{-12}$~eV, and the initial mass of each mode is $10^{-5}$ of the initial BH mass. 
    In total $10^3$ BHs are considered, distributed randomly in spin and log-uniformly in mass spanning from 3 $M_{\odot}$ to 100 $M_{\odot}$.
    The initial distribution is shown in panel (a).
    The distribution after 1 hour, 1 year, 10 years, $10^6$ years and $10^{10}$ years are shown in panels (b)-(f), respectively.
    In panels (b)-(f), the distribution of the previous moment is also shown in gray dots for comparison.
    In panels (b)-(f), the green curves from left to right correspond to the Regge trajectories of the dominant modes with $j=1,2,3,4,5$ from \eref{eq:a_critc}.}
    \label{fig:Regge_evo}
\end{figure*}

We first explain the mechanism behind the formation of the forbidden regions in the Regge plot with Fig.~\ref{fig:Regge_evo}. We consider a sample of $10^3$ BHs with spins randomly distributed from 0 to 1 and masses following a log-uniform distribution between 3 and 100 solar masses, together with a surrounding Proca field with vector mass $10^{-12}$~eV for each BH. Each vector mode begins with a mass of $10^{-5}$ times the initial BH mass. Numerical studies show that the subdominant modes and the GW emission have a very small effect on the Regge trajectories. Thus, we ignore their contributions in Eq.~\eqref{eq:Evoluton_ODE} and solve the evolution for each BH.

After an hour of evolution, the BHs with high spins and initial mass $M_0\in (15,50)M_\odot$ descend to the $^3S_{1,1}^0$ Regge trajectory, as illustrated in \figref{fig:Regge_evo}(b). BHs with high spin but smaller mass require more time to align with the Regge trajectory. BHs positioned below this trajectory do not experience the $j = 1$ stage, as the superradiant condition is not satisfied. After a year of evolution, shown in Fig.~\ref{fig:Regge_evo}(c), BHs with high spin and large mass further descend to the $^3P_{2,2}^0$ trajectory, while BHs with high spin and smaller mass just have enough time to arrive at the $^3S_{1,1}^0$ trajectory. The later evolution follows the same pattern. As time proceeds, trajectories with larger $j$ consistently emerge on the higher mass end and those with smaller $j$ gradually become irrelevant, which is shown in Figs.~\ref{fig:Regge_evo}(d)-(f).

As evolution progresses, the turning regions of different $j$ trajectories in the Regge plot shift downward and to the left, with new trajectories consistently emerging on the large-mass end. Meanwhile, the trajectories with the small $j$ values tend to vanish. Varying the vector mass leads to different Regge plots. Given the evolution time, the regions above the Regge trajectories are forbidden for BHs.

By comparing the observational data of BHs with the Regge plots obtained from different vector masses, we can constrain the mass of the vector field. Utilizing the 90 reported data from the LIGO-Virgo-KAGRA collaboration \cite{LIGOScientific:2018mvr,LIGOScientific:2020ibl,LIGOScientific:2021usb,LIGOScientific:2021djp}, with the corresponding parameter estimation samples from the online Gravitational-wave Transient Catalog,\footnote{\url{https://gwosc.org/eventapi/html/GWTC/}} we calculate the median for BH masses and spins. The sources with median mass below $3M_\odot$ are excluded since they may potentially be neutron stars.

\begin{figure}[ht]
    \centering
    \includegraphics[width=0.48\textwidth]{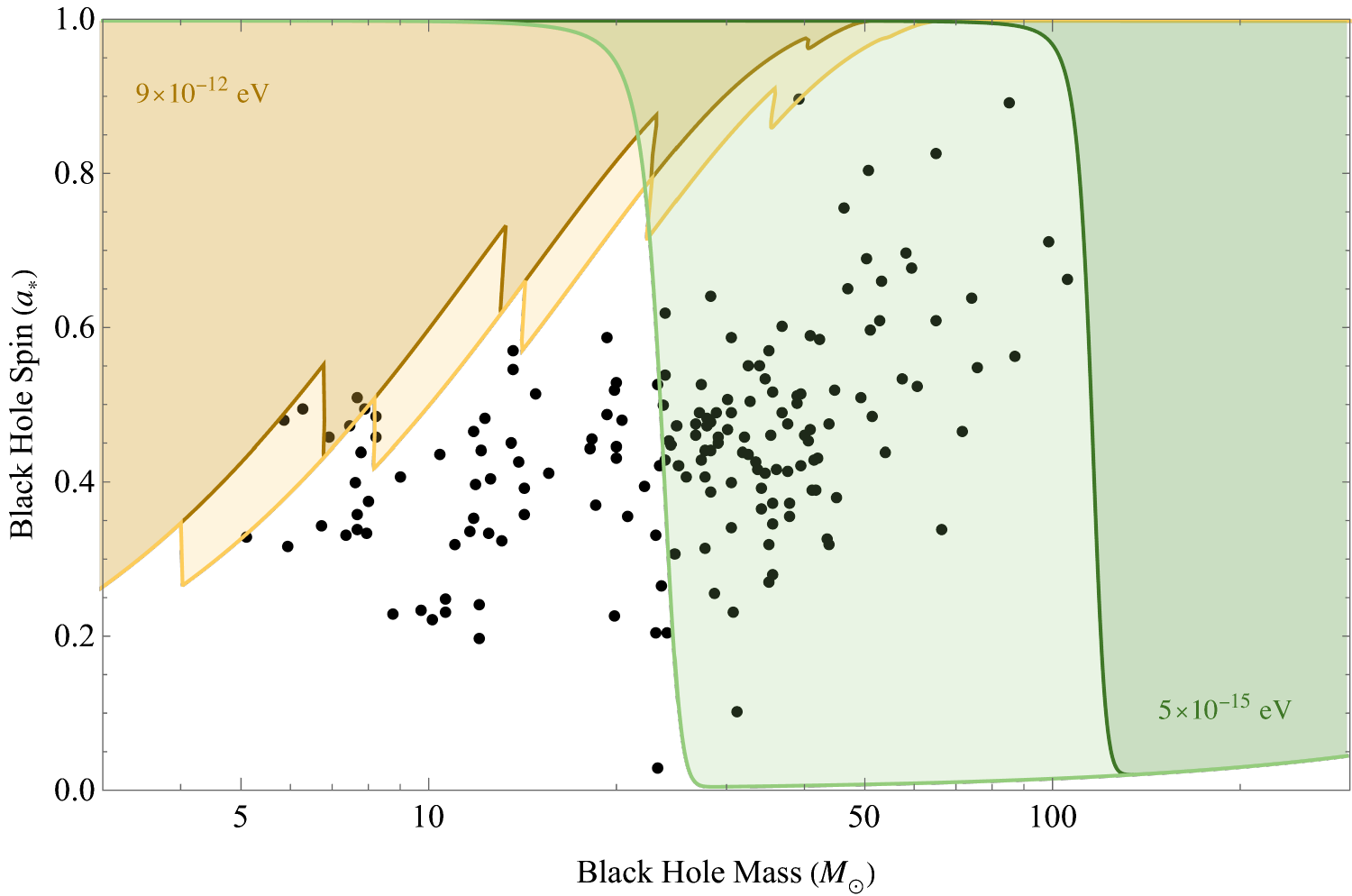}
    \caption{The comparison of the existing BH data and the Regge trajectories in the BH spin-mass distribution. Black points are the BH data from the confident GW events (see text for details). The zig-zag curves are Regge trajectories computed as in \figref{fig:Regge_evo} but with all initial BH spins set as 0.998, which provides conservative constraints. Two vector masses $9\times 10^{-12}$ eV (orange) and $5\times 10^{-15}$ eV (green) are plotted. For each mass, the darker and lighter curves correspond to $10^{6}$ and $10^{10}$ years evolution time, respectively. The shaded area above each curve is the forbidden region with the corresponding vector mass and the evolution time.}
    \label{fig:constraints}
\end{figure}

The resulting BH spin-mass distribution is plotted and compared with the Regge trajectories in \figref{fig:constraints}. The trajectories are computed similarly to Fig.~\ref{fig:Regge_evo}, but all initial BH spins are set as the limit value $a_{*\text{lim}}=0.998$ \cite{Thorne:1974ve}. We present two typical vector masses $9 \times 10^{-12}$ and $5\times 10^{-15}$~eV. For each mass, the darker color corresponds to $10^6$ years, while the lighter color represents $10^{10}$ years, approximately the age of the Universe. This range is given by binary BH formation models \cite{LIGOScientific:2016vpg,Arvanitaki:2016qwi}, and we conservatively shift the lower bound from $10^7$ years to $10^6$ years.

The shaded areas in \figref{fig:constraints} represent the forbidden regions.  By tuning the vector mass and repeating the calculation, we obtain a preliminary exclusion range of the vector mass,
\begin{align}
    5\times 10^{-15}\ \mathrm{eV}  <\mu< 9\times 10^{-12}\ \mathrm{eV}.
\end{align}
In this analysis, we have ignored the large uncertainties of the BH spin data. A strict treatment should include the data uncertainty and employ the Bayesian hierarchical method, as demonstrated in the scalar case studies \cite{Ng:2019jsx,Ng:2020ruv,Fernandez:2019qbj,Cheng:2022jsw}, which is beyond the scope of this work.

\section{Summary}
\label{sec:summary}
Ultralight bosons and a Kerr BH can form a BH-condensate system due to superradiant instability. In this work, we focus on the vector superradiance and carefully study the system's evolution. The evolution equations and the assumptions employed are given in Sec.~\ref{sec:sub_Evo_eqs}. Modes with different $j$ values govern different evolution stages and each $j$ stage can be further separated into three distinct phases: spin-down phase, and attractor phases for the subdominant and the dominant modes. We give explicit formulas to estimate the maximum masses for different modes, various timescales, and BH masses and spins. For direct GW observations, current and projected interferometers could potentially detect GWs emitted from the BH-vector-condensate systems. Indirectly, the GW events of the binary BHs mergers can be used to exclude vectors in the mass range $5\times 10^{-15}\ \mathrm{eV}  <\mu< 9\times 10^{-12}\ \mathrm{eV}$.

Our work could be extended by removing some of the assumptions shown in Sec.~\ref{sec:sub_Evo_eqs}. One promising avenue is to consider the impact of accretion, much like the scalar case \cite{Brito:2014wla,Guo:2022mpr}. By doing so, the life of each mode can still be split into different phases: spin-up, spin-down, attractor, and a possible quasi-normal phase depending on the accretion efficiency. Furthermore, the existence of companion stars could lead to a sharp depletion of the condensate, by causing resonance between growing and decaying modes \cite{Baumann:2018vus}. A third intriguing aspect to explore is the system's evolution in the presence of self-interaction, though the ghost instabilities may be large \cite{Clough:2022ygm,Mou:2022hqb}. Additionally, if higher-order analytical solutions are derived for the superradiance rate or if the numerical results are used, our work could be expanded to capture a broader region of mass coupling.

Below, we further summarize the main points covered in the previous sections. In Sec.~\ref{sec:vecBH}, we briefly review the properties of a free massive real vector in Kerr spacetime and the solutions of quasi-bound states. We give special attention to the hydrogen-like solutions that aid in calculating asymptotic GW emission fluxes. Subsequently, we discuss the two factors affecting the BH-condensate system: the superradiant instability and the GW emission. We first identify the dominant mode $^3(j-1)_{j,j}^0$ and the subdominant mode $^3(j-1)_{j,j}^1$. Then we calculate the GW emission fluxes in the limit of $M\mu\ll1$ following the method in Ref.~\cite{Brito:2014wla} with an additional flat-background-metric approximation, for two dominant modes ($^3S_{1,1}^0$ and $^3P_{2,2}^0$) and two subdominant modes ($^3S_{1,1}^1$ and $^3P_{2,2}^1$), with the interference effect accounted. Finally, we present the evolution equations for the system in Eqs.~\eqref{eq:Evoluton_ODE}, and list the assumptions we have employed.

In Sec.~\ref{sec:Single-mode}, we focus on the evolution with the fastest mode. Similar to the scalar case, this single-mode evolution also undergoes two main phases: the first one is governed by superradiance while the second one is controlled by GW emission. We give analytic expressions to estimate the characteristic quantities, which agree well with the numerical results in the $M\mu\ll1$ region. Then we discuss the effects of the initial parameters on the evolution.

In Sec.~\ref{sec:Multi-mode}, our attention shifts to the multi-mode evolution which is more realistic. The evolution can be segmented into stages with different $j$.  Each stage is further divided into three phases: spin-down phase, attractor phase for the subdominant mode, and attractor phase for the dominant mode. In the spin-down phase, superradiant modes grow exponentially while the BH loses its spin and mass. In the attractor phases, the BH spins align with the Regge trajectory of a specific mode, and all modes decay due to GW emission and/or absorption by the BH. The attractor for the subdominant mode is determined by solving \eref{eq:sub_attractor}, while the attractor for the dominant mode is given by \eref{eq:a_critc}. We provide estimates of the important quantities in the multi-mode evolution.

In Sec.~\ref{sec:detec}, we explore the direct and indirect detection approaches for vector superradiance, by studying the characteristic GW signals and by analyzing BH spin-mass distribution. For direct GW observations, we evaluate the characteristic strains of monochromatic waves for the $^3S_{1,1}^0$ mode and the $^3P_{2,2}^0$ mode, given in Eqs.~\eqref{eq:hc_0011} and \eqref{eq:hc_0122}, respectively. The strain of the GW beat is in \eref{eq:hc_beat} in the limit where the mass ratio of the subdominant mode to the dominant mode is much less than 1, i.e., $N_s/N_d\ll 1$. Notably, the beat modulation can reach $5\%$ even when this ratio is as small as $10^{-3}$. Comparing with the characteristic noise strain of the current and projected GW detectors, for the $j=m=1$ case, both the monochromatic GWs and the GW beats have good detectability for DECIGO (from $10^{-16}$ to $10^{-13}$ eV), advanced LIGO (from $10^{-13}$ to $10^{-12}$ eV) and space-based detectors (from $10^{-19}$ to $10^{-15}$ eV). For the $j=m=2$ case, DECIGO still has a very good potential but other detectors can only search for the sources in our neighborhood. For the indirect detection, we first explain the process underlying the formation of forbidden regions in the Regge plot.
By comparing the BH spin-mass data from the binary BH merger events to the forbidden regions with different vector masses, we found a preliminary exclusion range for the vector mass: $5\times 10^{-15}\ \mathrm{eV}  <\mu< 9\times 10^{-12}\ \mathrm{eV}$.

\section*{Acknowledgements}
We give special thanks to Dr. Gui-Rong Liang, for making the collaboration possible and his continuous support during the research. S.-S. B. and Y.-D. G., and H. Z. are supported by the National Natural Science Foundation of China (Grant No. 12075136) and the Natural Science Foundation of Shandong Province (Grant No. ZR2020MA094); N. J. is supported by the National Natural Science Foundation of China (Grants No. 12147163 and No. 12175099); X. Z. is supported by the National Natural Science Foundation of China (Grant No. 12473001), the National SKA Program of China (Grants No. 2022SKA0110200 and No. 2022SKA0110203), and the 111 Project (Grant No. B16009).
\\

\appendix

\section{The calculations of Regge trajectories}    

In this appendix, we compare our calculation of Regge trajectories with those reported in the literature. In Sec.~\ref{sec:indirect}, we obtain these trajectories by solving the evolution equations in Eqs.~\eqref{eq:Evoluton_ODE}, while most previous studies \cite{Arvanitaki:2010sy,Arvanitaki:2014wva,Arvanitaki:2016qwi,Ng:2019jsx,Ng:2020ruv,Fernandez:2019qbj,Cheng:2022jsw} adopt a different approach. They assume a BH undergoes a spin loss of $\Delta a_*\sim\mathcal{O}(1)$ and falls to the corresponding Regge trajectory as long as the evolution time $\tau$ is greater than  $90/\Gamma_{nljm}$. 

\begin{figure}[ht]
    \centering
    \includegraphics[width=0.48\textwidth]{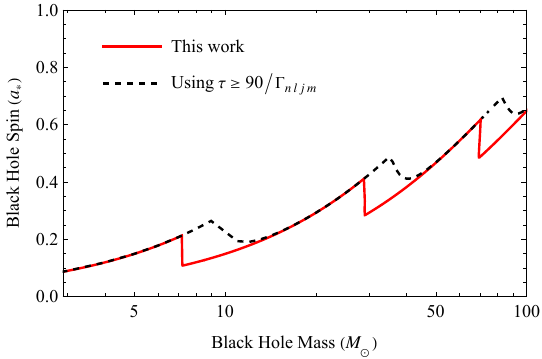}
    \caption{Comparison of the Regge trajectories obtained by solving the evolution equation to those with $\tau\geq 90/\Gamma_{nljm}$. In this plot, we choose the vector mass $10^{-12}$ eV, the initial BH spin 0.998, the initial condensate masses $10^{-5}M_0$, and the evolution time $10^6$~yr.}
    \label{fig:Regge_comparison}
\end{figure}

Fig.~\ref{fig:Regge_comparison} compares the results from both methods, using a vector mass of $10^{-12}$ eV, initial BH spin 0.998, and an evolution time of $10^6$ yr for illustration. Discrepancies exist in the connection regions of different Regge trajectories. The difference could be important if many observed BHs fall into these regions.

\end{document}